\documentclass[12pt,oneside,english]{book}
\usepackage{mathptmx}
\usepackage[T1]{fontenc}
\usepackage[latin9]{inputenc}
\usepackage{geometry}
\usepackage{array}
\usepackage{float}
\usepackage{amsmath}
\usepackage{graphicx}
\usepackage{setspace}
\usepackage{amssymb}

\makeatletter

\providecommand{\tabularnewline}{\\}

\renewcommand{\thepage}{}

\usepackage{babel}
\makeatother

\begin{document}

\renewcommand{\thepage}{}


\begin{titlepage}

\vfill \vfill \vfill \vfill

\begin{center}

\bigskip

\Huge

{\huge\bf Properties of Codes } \\

\bigskip

{\huge\bf in the Johnson Scheme} \\

\vfill \vfill

{\large Research Thesis}\\

\vfill \vfill \vfill

\normalsize

In Partial Fulfillment of the

 Requirements for the Degree of

 Master of Science in Applied Mathematics

\vfill \vfill \vfill

{\large \bf Natalia Silberstein}\\

\vfill \vfill

{\normalsize Submitted to the Senate of\\

the Technion - Israel Institute of Technology

}

\vfill \vfill \Large \Large \Large \Large \Large

\Large \Large \Large \Large

\vfill

{\normalsize Shvat 5767 \hfill Haifa \hfill February 2007}

\end{center}

\vfill

\end{titlepage}

\tableofcontents{}

\chapter*{Abstract}

\addcontentsline{toc}{chapter}{Abstract}

\pagenumbering{arabic}

Codes which attain the sphere packing bound are called perfect codes.
Perfect codes always draw the attention of coding theoreticians and
mathematicians. The most important metrics in coding theory on which
perfect codes are defined are the Hamming metric and the Johnson metric.
While for the Hamming metric all perfect codes over finite fields
are known, in the Johnson metric it was conjectured by Delsarte in
1970's that there are no nontrivial perfect codes. The general nonexistence
proof still remains the open problem.

Constant weight codes play an important role in various areas of coding
theory. They serve as building blocks for general codes in the Hamming
metric. One of the applications of constant weight codes is for obtaining
bounds on the sizes of unrestricted codes. In the same way as constant
weight codes play a role in obtaining bounds on the sizes of unrestricted
codes, doubly constant weight codes play an important role in obtaining
bounds on the sizes of constant weight codes .

In this work we examine constant weight codes as well as doubly constant
weight codes, and reduce the range of parameters in which perfect
codes may exist in both cases.

We start with the constant weight codes. We introduce an improvement
of Roos' bound for $1$-perfect codes, and present some new divisibility
conditions, which are based on the connection between perfect codes
in Johnson graph $J(n,w)$ and block designs. Next, we consider binomial
moments for perfect codes. We show which parameters can be excluded
for $1$-perfect codes. We examine $2$-perfect codes in $J(2w,w)$
and present necessary conditions for existence of such codes. We prove
that there are no $2$-perfect codes in $J(2w,w)$ with length less
then $2.5*10^{15}$.

Next we examine perfect doubly constant weight codes. We present properties
of such codes, that are similar to the properties of perfect codes
in Johnson graph. We present a family of parameters for codes whose
size of sphere divides the size of whole space. We then prove a bound
on length of such codes, similarly to Roos' bound for perfect codes
in Johnson graph.

Finally we describe Steiner systems and doubly Steiner systems, which
are strongly connected with the constant weight and doubly constant
weight codes respectively. We provide an anticode-based proof of a
bound on length of Steiner system, prove that doubly Steiner system
is a diameter perfect code and present a bound on length of doubly
Steiner system.

\chapter*{List of symbols and abbreviations}

\addcontentsline{toc}{chapter}{List of symbols and abbreviations}

\begin{tabular}{cl}
$\binom{n}{k}$ & binomial coefficient\tabularnewline
$S(r,v)$ & Stirling number of the second kind\tabularnewline
$GF(q)$ & Galois field of $q$ elements\tabularnewline
$N$ & set of coordinates\tabularnewline
$n$ & code length \tabularnewline
$w$ & code weight \tabularnewline
$d$ & code minimum distance \tabularnewline
$e$ & radius\tabularnewline
$C$ & code\tabularnewline
$J(n,w)$ & Johnson graph\tabularnewline
$\Phi_{e}(n,w)$ & size of a sphere of radius $e$ in $J(n,w)$\tabularnewline
$t-(n,w,\lambda)$ & $t$-design over $n$ elements and blocks of size $w$\tabularnewline
$S(t,w,n)$ & Steiner system over $n$ elements and blocks of size $w$\tabularnewline
$\varphi$ & code strength\tabularnewline
$S(t_{1},t_{2},w_{1},w_{2},n_{1},n_{2})$ & doubly Steiner system\tabularnewline
$\Phi_{e}(n_{1},n_{2},w_{1},w_{2})$ & size of a sphere of radius $e$ in doubly constant code\tabularnewline
\end{tabular}

\chapter{Introduction}

Codes which attain the sphere packing bound are called perfect code.
Perfect codes always draw the attention of coding theoreticians and
mathematicians. The most important metrics in coding theory on which
perfect codes are defined are the Hamming metric and the Johnson metric.

In the Hamming metric, all perfect codes over finite fields are known
\cite{MacW_and_Sloane}. They exist for only a small number of parameters,
while for other parameters their non-existence was proved \cite{Tiet,vanLint,Zinoviev,MacW_and_Sloane}.
The nonexistence proof is based on Lloyd's polynomials. No nontrivial
perfect code is known over other alphabets and for most parameters
it was proved that they do not exist \cite{Best}.

As for the Johnson metric, it was conjectured by Delsarte \cite{Delsarte}
in 1973 that there are no nontrivial perfect codes. Many attempts
were made during the last 35 years to prove this conjecture. These
attempts used Lloyd polynomials, anticodes, designs and number theory.
However, the previous research yielded only partial results and the
general nonexistence is yet to be proved.

Perfect codes in the Johnson metric have a strong connection to constant
weight codes.

Constant weight codes play an important role in various areas of coding
theory. One of their applications is in obtaining lower and upper
bounds on the sizes of unrestricted codes for given length and minimum
Hamming distance \cite{Johnson,MacW_and_Sloane}.

In the same way as constant weight codes are used for obtaining bounds
on the sizes of unrestricted codes, doubly constant weight codes play
an important role in obtaining bounds on the sizes of constant weight
codes \cite{Agrell}. A natural question is whether there exist perfect
doubly constant weight codes.

\section{Definitions}

A \emph{binary unrestricted code} of length $n$ is the set of binary
words of length $n$.

The \emph{weigh}t of a word is the number of ones in the word.

A \emph{constant weight cod}e of length $n$ and weight $w$ is a
binary code whose codewords have constant weight $w.$

A \emph{doubly constant weight code} of length $n$ and weight $w$
is a constant weight code of length $n$ and weight $w$, with $w_{1}$
ones in the first $n_{1}$ positions and $w_{2}$ ones in the last
$n_{2}$ positions, where $n=n_{1}+n_{2}$ and $w=w_{1}+w_{2}$ .

The \emph{Hamming distance} (or H-distance in short) between two words
of the same length $n$ is the number of coordinates in which they
differ.

If we define the distance between two words, $x$ and $y$ of the
same weight $w$ and the same length $n$, as half their H-distance,
we obtain a new metric which is called the \emph{Johnson metric} and
the distance is called the \emph{Johnson distance} (or J-distance
in short).

Let $A(n,d)$ denote the maximum number of codewords in a binary code
of length $n$ and minimum H-distance $d$

Let $A(n,d,w)$ denote the maximum number of codewords in a constant
weight code of length $n$, weight $w$ and minimum H-distance $d$.

A $(w_{1},n_{1},w_{2},n_{2},d)$ code is a doubly constant weight
code with $w_{1}$ ones in the first $n_{1}$ positions and $w_{2}$
ones in the last $n_{2}$ positions, and minimum J-distance $d$ .

Let $T(w_{1},n_{1},w_{2},n_{2},\delta)$ denote the maximum number
of codewords in a $(w_{1},n_{1},$$w_{2},$ $n_{2},d)$ code, where
$\delta=2d$ is a H-distance.

\subsection{Block designs}

\noindent There is a tight connection between constant weight codes
and block designs.

In the next chapters we will use the following terminology and properties
of block designs.

\noindent \textbf{Definition}. Let $t,n,w,\lambda$ be integers with
$n>w\geq t$ and $\lambda>0$. Let $N$ be an $n$-set (i.e. a set
with $n$ elements), whose elements are called points or sometimes
(for historical reasons) varieties. A $t-(n,w,\lambda)$ \emph{design}
is a collection $C$ of distinct $w$- subsets called blocks of $N$
with the property that any $t$-subset of $N$ is contained in exactly
$\lambda$ blocks of $C$.

\noindent \textbf{Example}. If we take the lines as blocks, the seven
points and seven lines (one of which is curved) of Figure \ref{cap:2-(7,3,1)-design}
form a $2-(7,3,1)$ design, since there is a unique line trough any
two of the seven points. The seven blocks are \[
013,\;124,\;235,\;346,\;450,\;561,\;602.\]

\begin{figure}[H]
\caption{\label{cap:2-(7,3,1)-design}2-(7,3,1) design}

\begin{centering}
\includegraphics{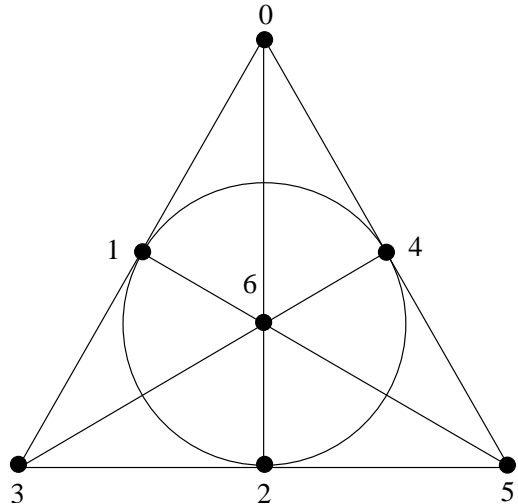}
\par\end{centering}
\end{figure}

The following two theorems are well known (see \cite{MacW_and_Sloane}
for reference).

\noindent \textbf{Theorem 1}. If $s<t$ then every $t$-design is
also an $s-$design.

\begin{description}
\item [{Notes}]~
\end{description}
\begin{enumerate}
\item In a $t-(n,w,\lambda)$ design the total number of blocks is \[
b=\lambda\frac{\binom{n}{t}}{\binom{w}{t}}\]

\item The existence of a $t-(n,w,\lambda)$ design implies the existence
of $(t-1)-(n-1,w-1,\lambda)$ design (called the derived design) and
$(t-1)-(n,w,\lambda')$ design, and hence it must satisfy certain
divisibility conditions:
\end{enumerate}
\textbf{Theorem 2.} A necessary condition for a $t-(n,w,\lambda)$
design to exist, is that the numbers \[
\lambda\frac{\binom{n-i}{t-i}}{\binom{w-i}{t-i}}\]
must be must be integers , for $0\leq i\leq t$.

A Steiner system is simply a $t-$design with $\lambda=1$.

\noindent \textbf{Definition}. A \emph{Steiner system} $S(t,w,n)$
is a collection of $w-$subsets (blocks) of $n$-set $N$ such that
every $t$- subset of $N$ is contained in exactly one of the blocks.

Note that we use $S(t,w,n)$ as an equivalent of $t-(n,w,1)$. Thus
the example of Figure \ref{cap:2-(7,3,1)-design} is an $S(2,3,7)$.

\noindent \textbf{Corollary 3}. A Steiner system $S(t,w,n)$ has $\binom{n}{t}/\binom{w}{t}$
blocks.

\noindent \textbf{Corollary} 4. If there exists a Steiner system $S(t,w,n)$
for $t\geq1$, then there exists a Steiner system $S(t-1,w-1,n-1)$.

\noindent \textbf{Corollary} 5. A necessary condition for a Steiner
system $S(t,w,n)$ to exist, is that the numbers $\binom{n-i}{t-i}/\binom{w-i}{t-i}$
must be integers, for $0\leq i\leq t$.

\noindent \textbf{Incidence Matrix}. Given a $t-(n,w,\lambda)$ design
with $n$ points $P_{1},...,P_{n}$ and $b$ blocks $B_{1},...,B_{b}$
its $b\times n$ \emph{incidence} \emph{matrix} $A=(a_{ij})$ is defined
by\[
a_{ij}=\left\{ \begin{array}{c}
1\;\;\mbox{if }P_{j}\in B_{i}\\
0\;\;\mbox{if }P_{j}\notin B_{i}\end{array}\right.\]

\noindent For example the incidence matrix of the design of Figure
\ref{cap:2-(7,3,1)-design} is

\[
A=\left(\begin{array}{ccccccc}
1 & 1 & 0 & 1 & 0 & 0 & 0\\
0 & 1 & 1 & 0 & 1 & 0 & 0\\
0 & 0 & 1 & 1 & 0 & 1 & 0\\
0 & 0 & 0 & 1 & 1 & 0 & 1\\
1 & 0 & 0 & 0 & 1 & 1 & 0\\
0 & 1 & 0 & 0 & 0 & 1 & 1\\
1 & 0 & 1 & 0 & 0 & 0 & 1\end{array}\right).\]

\noindent \textbf{Codes and Designs.} To every block in a $t-(n,w,\lambda)$
design corresponds a row of the incidence matrix $A$. If we think
of these rows as codewords, the $t$-design forms a constant weight
code $C$ of length $n$ and weight $w$.

The largest $t$ of a code $C$ for which the code is a $t$-design
is called the \emph{strength} of the code.

\section{Perfect codes in the Hamming metric}

A code $C$ of length $n$ and minimum H-distance $d=2e+1$ is called
an \emph{$e$-perfect} if for each vector $v$ of length $n$ there
exists a unique element $c\in C$, such that the H-distance between
$v$ and $c$ is at most $e$.

There are the \emph{trivial} perfect codes: a code containing just
one codeword, or the whole space, or a binary repetition code of odd
length.

Three types of perfect codes in Hamming metric were discovered in
the late 1940's:

\begin{enumerate}
\item The linear single-error-correcting Hamming codes $\left[n=\frac{q^{m}-1}{q-1},n-m,3\right]$,
\item The binary $\left[23,12,7\right]$ Golay code
\item The ternary $\left[11,6,5\right]$ Golay code
\end{enumerate}
\textbf{Theorem 6} \cite{Tiet,vanLint} A nontrivial perfect code
over any field $GF(q)$ must have the same parameters as one of the
Hamming or Golay codes.

For non-field alphabets only trivial codes are known and it was proved
that for most other parameters they do not exist. \cite{Best}

\section{Perfect codes in the Johnson metric (survey of known results)}

We associate the \emph{Johnson graph} $J(n,w)$ with the Johnson space
for given positive integers $n$ and $w$ such that $0\leq w\leq n$.
The vertex set $V_{w}^{n}$ of the Johnson graph consists of all $w$-subsets
of a fixed $n$-set $N=\left\{ 1,2,...,n\right\} $. Two such $w$-subsets
are adjacent if and only if their intersection is of size $w-1$.
A code $C$ of such $w$-subsets is called an $e$-\emph{perfect code}
in $J(n,w)$ if the $e$-spheres with centers at the codewords of
$C$ form a partition of $V_{w}^{n}$. In other words, $C$ is an
$e$-perfect code if for each element $v\in V_{w}^{n}$ there exists
a unique element $c\in C$ such that the distance between $v$ and
$c$ is at most $e$.

A code $C$ in $J(n,w)$ can be described as a collection of $w$-subsets
of $N$, but it can be also described as a binary code of length $n$
and constant weight $w$. From a $w$-subset $S$ we construct a binary
vector of length $n$ and weight $w$ with ones in the positions of
$S$ and zeros in the positions of $N\setminus S$. In the sequel
we will use a mixed language of sets and binary vectors.

There are some \emph{trivial perfect codes} in $J(n,w)$:

\begin{enumerate}
\item $V_{w}^{n}$ is $0$-perfect.
\item Any $\left\{ v\right\} $, $v\in V_{w}^{n}$, $w\leq n-w$, is $w$-perfect.
\item If $n=2w$, w odd, any pair of disjoint $w$-subsets is $e$-perfect
with $e=\frac{1}{2}(w-1)$.
\end{enumerate}
\noindent Delsarte conjectured that there are no perfect codes in
$J(n,w)$, except for these trivial perfect codes. In his seminal
work from 1973 \cite{Delsarte} , he wrote :

{}``After having recalled that there are {}``very few'' perfect
codes in the Hamming schemes, one must say that, for $1<\delta<n$,
there is not a single one known in the Johnson schemes. It is tempting
to risk the conjecture that such codes do not exist. {}``

Indeed, Delsarte omitted the trivial perfect codes (we will omit them
too, unless otherwise stated, so when we say perfect codes we mean
nontrivial perfect codes), and his conjecture on the nonexistence
of perfect codes in the Johnson spaces has provided plenty of ground
for research in the years which followed. Due to the fact that in
the Hamming spaces over $GF(q)$ all parameters for which perfect
codes exist were known, special emphasis was given to the Johnson
spaces. However, not many significant results were produced.

A connected graph $\Gamma$ with diameter $d$ is called \emph{distance-regular}
if for any vertices $x$ and $y$ of $\Gamma$ and any integers $0\leq i$,
$j\leq d$ , the number of vertices $z$ at distance $i$ from $x$
and at distance $j$ from $y$ depends only on $i$, $j$ and $k:=\mbox{dis}\mbox{t}(x,y)$
and not on the choice of $x$ and $y$ themselves.

The following theorem is due to Delsarte \cite{Delsarte}:

\noindent \textbf{Theorem 7} : Let $X$ and $Y$ be subsets of the
vertex set $V$ of a distance regular graph $\Gamma$, such that nonzero
distances occurring between vertex in $X$ do not occur between vertices
of $Y$. Then $\mid X\mid\cdot\mid Y\mid\leq\mid V\mid$.

A subset $X$ of $V$ is called an \emph{anticode} with diameter \textbf{$D$},
if $D$ is the maximum distance occurring between vertices of $X$.

Anticodes with diameter $D$ having maximal size are called \emph{optimal
anticodes}\textbf{\emph{.}}

\noindent Let $\Gamma$ be a connected graph. We denote by $d_{\Gamma}(x,y)$
the length of the shortest path from $x$ to $y$. $\Gamma$ is said
to be \emph{distance} \emph{transitive} if, whenever $x,x',y,y'$
are vertices with $d_{\Gamma}(x,x')$=$d_{\Gamma}(y,y')$, there is
an automorphism $\gamma$ of $\Gamma$ with $\gamma(x)=y$ and $\gamma(x')=y'$
. A distance-transitive graph is obviously distance regular.

Biggs \cite{Biggs} showed that the natural setting for the existence
problem of perfect codes is the class of distance transitive graphs.
Biggs claims that the class of distance transitive graphs includes
all interesting schemes, such as the Hamming scheme and the Johnson
scheme, and developed a general theory and a criterion for the existence
of perfect codes in a distance-transitive graph. He showed that this
criterion implies Lloyd's theorem, which is used in the Hamming scheme
to prove the nonexistence of perfect codes in all cases.

Bannai \cite{Bannai} proved the nonexistence of $e$-perfect codes
in $J(2w-1,w)$ and $J(2w+1,w)$, for $e\geq2$. He used an analogue
to Lloyd's theorem and some number-theoretic results.

Hammond \cite{Hammond} extended this result and showed that $J(n,w)$
can not contain a nontrivial perfect code for $n\in\left\{ 2w-2,2w-1,2w+1,2w+2\right\} $.

\noindent \textbf{Theorem 8} \cite{Hammond}. There are no perfect
codes in $J(2w-2,w)$, $J(2w-1,w)$, $J(2w+1,w)$ and $J(2w+2,w)$.

However, the most significant result, in the first twenty years following
Delsarte's conjecture, was given in 1983 by Roos \cite{Roos}.

\noindent \textbf{Theorem 9} \cite{Roos}. If an $e$-perfect code
in $J(n,w)$, $n\geq2w$, exists, then $n\leq(w-1)\frac{2e+1}{e}$.

The proof of Roos was based on anticodes. By using Theorem 7, Roos
noticed that if an $e$-perfect code exists, then the $e$-spheres
should be optimal anticodes with diameter 2e. He proceeded to find
anticodes in $J(n,w)$ and obtained his result by comparing them to
the $e$-spheres.

Etzion in \cite{on_perfect} give a different simple proof of this
theorem and in \cite{Schwartz} Etzion and Schwartz show that no nontrivial
$e$-perfect code achieves Roos' bound with equality.

Another approach was shown by Etzion in \cite{nonexist}. He proved
that if there exists a nontrivial $e$-perfect code $C$ in $J(n,w)$,
then many Steiner systems are embedded in $C$. Using Etzion's approach,
the necessary conditions for the existence of Steiner systems imply
necessary conditions for the existence of perfect codes in the Johnson
graph. Moreover, Etzion developed a new concept called configuration
distribution, which is akin to the concept of weight distribution
for codes in the Hamming metric. Using this concept, combined with
the necessary conditions derived from Steiner systems, many parameters
were found, for which $e$-perfect codes do not exists in $J(n,w)$.
We summarize the main results given in \cite{nonexist,on_perfect}:

\noindent \textbf{Lemma 10}. If $C$ is an $e$-perfect code in the
Johnson scheme then its minimum H-distance is $4e+2$.

\noindent \textbf{Lemma 11.} If $C$ is an $e$-perfect code in the
$J(n,w)$ then $A(n,4e+2,w)=\left|C\right|$.

Let $N=\left\{ 1,2,...,n\right\} $ be the $n$-set. From a Steiner
system $S(t,w,n)$ we construct a constant-weight code on $n$ coordinates
as follows. From each block $B$ we construct a codeword with ones
in the positions of $B$ and zeros in the positions of $N\setminus B$.
This construction leads to the following well known theorem \cite{optStS}.

\noindent \textbf{Theorem 12.} $A(n,2(k-t+1),k)=\frac{n(n-1)\cdot\cdot\cdot(n-t+1)}{k(k-1)\cdot\cdot\cdot(k-t+1)}$
if and only if a Steiner system $S(t,k,n)$ exists.

From Theorem 12 and Lemma 10 we immediately infer the following result.

\noindent \textbf{Lemma 13.} If $C$ is an $e$-perfect code in $J(n,w)$
which is also a Steiner system, then it is a Steiner system $S(w-2e,w,n)$.

The next lemma is a simple observation of considerable use.

\noindent \textbf{Lemma 14}. The complement of an $e$-perfect code
in $J(n,w)$ is an $e$-perfect code in $J(n,n-w)$.

If we combine Lemma 4 with the fact that the J-distance between words
of an $e$-perfect code is at least $2e+1$, we get:

\noindent \textbf{Corollary 15}. If an $e$-perfect code exists in
$J(n,w)$, then $w\geq2e+1$ and $n-w\geq2e+1$.

For a given partition of $N$ into two subsets, $A$ and $B$, such
that $\left|A\right|=k$ and $\left|B\right|=n-k$, let \emph{configuration}
$(i,j)$ consist of all vectors with weight $i$ in the positions
of $A$ and weight $j$ in the positions of $B$.

For an $e$-perfect code $C$ in $J(n,w)$, we say that $u\in C$
\emph{$J$-covers} $v\in V_{w}^{n}$ if the J-distance between $u$
and $v$ is less than or equal to $e$. For a given two subsets $u$
and $v$ we say that $u$ \emph{$C$-covers} $v$ if $v$ is a subset
of $u$.

\noindent \textbf{Theorem 16.} If an $e$-perfect code exists in $J(n,w)$
, then a Steiner system $S(e+1,2e+1,w)$ and a Steiner system $S(e+1,2e+1,n-w)$
exist.

\noindent \textbf{Theorem 17}. If an $e$-perfect code exists in $J(n,w)$,
then a Steiner system $S(2,e+2,w-e+1)$ and a Steiner system $S(2,e+2,n-w+e-1)$
exist.

\noindent \textbf{Corollary 18}. If an $e$-perfect code exists in
$J(n,w)$, then $n-w\equiv w\equiv e(\mbox{mod }e+1)$ and hence $e+1$
divides $n-2w$.

\noindent \textbf{Theorem 19}. Except for the Steiner systems $S(1,w,n)$
and $S(w,w,n)$, there are no more Steiner systems which are also
perfect codes in the Johnson scheme.

\noindent \textbf{Theorem 20}. An $e$-perfect code in $J(2w,w)$
is self-complement, i.e., the complement of the code is equal to the
code.

\noindent \textbf{Theorem 21}. There are no $e$-perfect codes in
$J(2w+p,w)$, $p$ prime, in $J(2w+2p,w)$, $p$ is a prime, $p\neq3$,
and in $J(2w+3p,w)$, $p$ is a prime, $p\neq2,3,5$.

\noindent \textbf{Theorem 22}. If an $e$-perfect code exists in $J(n,w)$and
$n<(w-1)(2e+1)/e$, then a $S(2,e+2,n-w+2)$ exists.

\noindent \textbf{Corollary 23}. If an $e$-perfect code in $J(n,w)$
exists and $w\leq n-w$, then a $S(2,e+2,w+2)$ exists.

Now, we consider the Steiner systems which are embedded in an $e$-perfect
code in $J(n,w)$. By using the necessary condition for existence
of Steiner system, we have the following results.

\noindent \textbf{Theorem 24}. Assume there exists an $e$-perfect
code in $J(n,w)$.

\begin{itemize}
\item If $e$ is odd then $n$ is even and $(e+1)(e+2)$ divides $n-2w$.
\item If $e$ is even and $n$ is even then$(e+1)(e+2)$ divides $n-2w$.
\item If $e$ is even and $n$ is odd then $e\equiv0(\mbox{mod }4)$ and
$\frac{(e+1)(e+2)}{2}$ divides $n-2w$.
\end{itemize}
\textbf{Corollary 25}. There are no perfect codes in:

\begin{itemize}
\item $J(2w+p^{i},w)$, $p$ is a prime and $i\geq1$.
\item $J(2w+pq,w)$, $p$ and $q$ primes, $q<p$, and $p\neq2q-1$.
\end{itemize}
Etzion and Schwartz \cite{Schwartz} introduced the concept of $t$-regular
codes.

We summarize some of the relevant results from \cite{Schwartz}.

\noindent \textbf{Theorem 26.} If an $e$-perfect code $C$ in $J(n,w)$
is $t$-regular, then\[
\Phi_{e}(n,w)\left|\binom{n-i}{w-i}\right.,\]
for all $0\leq i\leq t$, where $\Phi_{e}(n,w)$ denotes the size
of sphere of radius $e$.

\noindent Define the following polynomial :

\[
\sigma_{e}(w,a,t)=\sum_{j=0}^{e}(-1)^{j}\binom{t}{j}\sum_{i=0}^{e-j}\binom{w-j}{i}\binom{w+a-t+j}{i+j}.\]

\noindent \textbf{Theorem 27.} Let $C$ be an $e$-perfect code in
$J(2w+a,w),$ and let $1\leq t\leq w$. If $\sigma_{e}(w,a,m)\neq0$
for all the integers $1\leq m\leq t$, then $C$ is $t$-regular.

\noindent \textbf{Theorem 28}. If a $1$-perfect code exists in $J(2w+a,w)$,
then it is $t$-regular for all \begin{equation}
0\leq t\leq\frac{2w+a+1-\sqrt{(a+1)^{2}+4(w-1)}}{2}.\label{eq:18}\end{equation}

\noindent \textbf{Theorem 29.} There are no $1$-perfect codes in
$J(n,w)$, when \[
\Phi_{1}(n,w)=1+w(n-w)\equiv0(\mbox{mod }4).\]

\noindent \textbf{Theorem 30.} If an $e$-perfect code, $e\geq2$,
exists in $J(2w+a,w)$, then it is $t$-regular for all $0\leq t\leq\frac{w}{e}-e$.

\noindent \textbf{Corollary 31}. If an $e$-perfect code exists in
$J(n,w)$, then it is $e$-regular.

\noindent \textbf{Theorem 32.} For all $e\geq2$, there exists $W_{e}>0$,
such that for all $w\geq W_{e}$, all $e$-perfect codes in $J(2w+a,w)$
are $\left\lfloor \frac{w}{2}\right\rfloor $-regular.

\noindent \textbf{Theorem 33.} There are no $e$-perfect codes in
$J(n,w)$, $e\geq2$, which are also $\left\lfloor \frac{w}{2}\right\rfloor $-regular,
when $\Phi_{e}(n,w)\equiv0(\mbox{mod }p^{2})$, $p$ a prime.

\noindent \textbf{Theorem 34.} Let $p$ be a prime, and $e\equiv-1(\mbox{mod }p^{2})$.
If an $e$-perfect code exists in $J(n,w)$, then \[
\Phi_{e}(n,w)\equiv0(\mbox{mod }p^{2}).\]

\noindent \textbf{Corollary 35}. For any given $e\geq2$ , $e\equiv-1(\mbox{mod }p^{2})$,
$p$ a prime, there are finitely many nontrivial $e$-perfect codes
in the Johnson graph.

\noindent \textbf{Theorem 36}. There are no nontrivial $3$-perfect
, $7$-perfect, $8$-perfect codes in the Johnson graph.

Martin \cite{Martin} also examined the existence problem when he
considered completely-regular subsets in his thesis. He found that
if $e=1$, then perfect codes must obey some numerical formula: $w=rs+1$
and $n=2rs+r-s+1$. Etzion \cite{conf_dist} has shown that these
observations are implied from (\ref{eq:18}) .

Ahlswede, Aydinian and Khachatrian \cite{Ahlswede} gave a new interesting
definition of \emph{diameter-perfect codes} (D-perfect codes). They
examined a variant of Theorem 7(of Delsarte). Let $\Gamma$ be a distance-regular
graph with a vertex set $V$. If $A$ is an anticode in $\Gamma$,
denote by $D(A)$ the diameter of $A$. Now let \[
A^{*}(D)=\mbox{max }\left\{ \left|A\right|\;:\; D(A)\leq D\right\} .\]

\noindent \textbf{Theorem 37.} If $C$ is a code in $\Gamma$ with
minimum distance $D+1$, then $\left|C\right|\leq\left|V\right|A^{*}(D)^{-1}$.

They continued with the following new definition for perfect codes.
A code $C$ with minimum distance $D+1$ is called $D$-perfect if
Theorem 37 holds with equality. This is a generalization of the usual
definition of $e$-perfect codes as $e$-spheres are anticodes with
diameter $2e$.

Gordon \cite{Gordon} proved that size of sphere of $1$-perfect code
in $J(n,w)$ is squarefree, and for each prime $p_{i}\left|\Phi_{1}(n,w)\right.$,
there is an integer $\alpha_{i}$ such that $p_{i}^{\alpha_{i}}$
must be close to $n-w$, moreover, the $\alpha_{i}$'s are distinct
and pairwise coprime, and the sum of their reciprocals is close to
two.

\section{Organization of this work}

The rest of this thesis is organized as follows.

In Chapter 2 we examine perfect codes in the Johnson graph. We start
by a brief survey of the techniques concerning the existence of perfect
codes in the Johnson graph, which are relevant to our work. Then we
introduce the improvement of Roos bound for $1$-perfect codes, and
present some new divisibility conditions. Next, we consider binomial
moments for perfect codes and show which general parameters can be
ruled out. Finally we examine $2$-perfect codes in $J(2w,w)$ and
present necessary conditions for existence of such codes, using Pell
equations.

In Chapter 3 we examine perfect doubly constant weight codes. We present
the properties of such codes, that are similar to the properties of
perfect codes in Johnson graph, construct the family of parameters
for codes whose sphere divides the size of whole space and finally
prove the bound on length on such codes, that is similar to Roos'
bound for perfect codes in Johnson graph.

Chapter 4 deals with Steiner systems and doubly Steiner systems. We
provide an anticode-based proof of the bound on Steiner system, prove
that doubly Steiner system is a diameter perfect code and present
the bound on the size of doubly Steiner system.

\chapter{Perfect codes in $J(n,w)$}

\section{$t$-designs and codes in $J(n,w)$}

In this section we use $t$ - designs and and the strength of the
code for excluding Johnson graphs in which there are no $e$-perfect
codes. We introduce the notion of $t\mbox{-regular }$codes, and their
properties, as presented in \cite{Schwartz}.

In $J(n,w)$, let \[
\Phi_{e}(n,w)=\sum_{i=0}^{e}\binom{w}{i}\binom{n-w}{i},\]

\noindent denote size of sphere of radius $e$. The number of codewords
in an $e$-perfect code $C$ in $J(n,w)$ is \[
|C|=\frac{\binom{n}{w}}{\Phi_{e}(n,w)}\]
by the sphere packing bound, hence \[
\Phi_{e}(n,w)\left|\binom{n}{w}.\right.\]

However, we learn much more about perfect codes, by using the approach
which was presented in \cite{Schwartz}. Now we introduce the definition
of $t$-regular codes:

\noindent \textbf{Definition 1.} Let $C$ be a code in $J(n,w)$ and
let $A$ be a subset of the coordinate set $N$. For $0\leq i\leq\left|A\right|$
we define \[
C_{A}(i)=\left|\left\{ c\in C\;:\left|c\cap A\right|=i\right\} \right|.\]
Also, for each $I\subseteq A$ we define \[
C_{A}(I)=\left|\left\{ c\in C\;:\; c\cap A=I\right\} \right|.\]

\noindent \textbf{Definition} 2. A code $C$ in $J(n,w)$ is said
to be $t$-regular, if the following two conditions hold:

(c.1) There exist numbers $\alpha(0),...,\alpha(t)$ such that if
$A\subset N$, $\left|A\right|=t$, then $C_{A}(i)=\alpha(i)$ for
all $0\leq i\leq t$.

(c.2) For any given $t$-subset $A$ of $N$, there exist numbers
$\beta_{A}(0),...,\beta_{A}(t)$ such that if $I\subset A$ then $C_{A}(I)=\beta_{A}(\left|I\right|)$.

\noindent Note that if a code is $t$ -regular, $t\geq1$, then it
is also $(t-1)$- regular.

It was proved in \cite{conf_dist} that a code $C$ in $J(n,w)$ is
$t$-regular if and only if it forms $t$-design. The strength of
an $e$-perfect code $C$ can be used to exclude the existence of
perfect codes by the following theorem \cite{Schwartz}.

\noindent \textbf{Theorem 37}. If an $e$-perfect code $C$ in $J(n,w)$
is $t$-regular, then\[
\Phi_{e}(n,w)\left|\binom{n-i}{w-i},\right.\]
for all $0\leq i\leq t$.

It was proved in \cite{Schwartz} that if $C$ is an $e$-perfect
code in $J(n,w)$ with strength $\varphi$ then\[
\sum_{i=0}^{e}(-1)^{i}\binom{\varphi+1}{i}\sum_{j=0}^{e-i}\binom{w-i}{j}\binom{n-w-\varphi-1+i}{i+j}=0\]

\noindent and for $t\leq\varphi$\[
\sum_{i=0}^{e}(-1)^{i}\binom{t}{i}\sum_{j=0}^{e-i}\binom{w-i}{j}\binom{n-w-t+i}{i+j}\neq0.\]
Therefore, the polynomial $\sigma_{e}(n,w,t)=\sum_{i=0}^{e}(-1)^{i}\binom{t}{i}\sum_{j=0}^{e-i}\binom{w-i}{j}\binom{n-w-t+i}{i+j}$,
defined in \cite{Schwartz} satisfies the following condition: the
smallest positive integer $\varphi$ for which $\sigma_{e}(n,w,\varphi+1)=0$
is the strength of $C$.

When $e=1$, $\sigma_{e}(n,w,t)$ is quadratic equation and $\varphi$
is easily computed:\begin{equation}
\varphi=\frac{n-1-\sqrt{(n-2w+1)^{2}+4(w-1)}}{2}\label{eq:1}\end{equation}
Note, that when $e\geq2$, $\sigma_{e}(n,w,t)$ is much more complicated
polynomial, and it is tempting to conjecture that there are no integer
solutions to $\sigma_{e}(n,w,t)=0$ for $e>2$ .

\subsection{Divisibility conditions for $1$-perfect codes in $J(n,w)$}

Now we prove the theorem which provides divisibility conditions for
$1$-perfect codes in $J(2w+a,w)$.

\noindent \textbf{Theorem 38.} If there exists a $1$-perfect code
$C$ with strength $w-d$ for some $d\geq0$ in $J(2w+a,w)$ , then

\begin{enumerate}
\item $w-d$ $\equiv0,1,4\mbox{ or }9(\mbox{mod }12)$
\item $\lambda:=\frac{\prod_{i=0}^{d-2}(wd-(d+i(d-1)))}{(d-1)!(d-1)^{d-1}d(w-d+1)}\in\mathbb{Z}$
\item $\lambda\prod_{j=1}^{s}$$\left[\frac{wd+jd-(j+1)}{(d-1)(d+j)}\right]\in\mathbb{Z},\:0\leq s\leq w-d$
\end{enumerate}
\emph{Proof}\textbf{.} Assume that there exists a $1$-perfect code
in $J(2w+a,w).$ Therefore, by (\ref{eq:1}), the strength of $C$
is \[
\frac{2w+a-1-\sqrt{(a+1)^{2}+4(w-1)}}{2}.\]
Define the following function of $w$ and $a$\[
f(w,a)=\frac{2w+a-1-\sqrt{(a+1)^{2}+4(w-1)}}{2}.\]
Note that $f(w,a)$ is an increasing function of $a$.

\noindent Now suppose that $f(w,a)=w-d$. Therefore, we get the following
expression for $a$:

\[
a=\frac{w-d^{2}+d-1}{d-1},\]
therefore, $d>1$.

Now we use the following lemma \cite{Schwartz}:

\noindent \textbf{Lemma 39.} If there exists a $1$-perfect code in
$J(n,w)$ then either $w\equiv n-w\equiv1(\mbox{mod }12)$ or $w\equiv n-w\equiv7(\mbox{mod }12)$.

In particular, $w\equiv1(\mbox{mod }6)$ and $\left.6\right|a$ ,
hence given that $w=6k+1$ for some integer $k$, it follows that
\[
6\left|a=\frac{6k-d^{2}+d}{d-1}\right.\]
or \[
6\left|d^{2}-d.\right.\]
Therefore, $d\equiv0(\mbox{mod }3)$ or $d\equiv1(\mbox{mod }3)$.
We write this result modulo $12$: $d\equiv0,1,3,4,6,7,9\mbox{ or }10(\mbox{mod }12).$

Now we consider all the values of $d$ modulo 12 and relate them to
the values of $w$ and $w-d$, e.g. the strength, modulo 12.

\noindent Since $a=\frac{w-d^{2}+d-1}{d-1}$, \[
w=(d-1)a+d^{2}-d+1\]
From Lemma 39, $\left.12\right|a$, thus\[
w\equiv d^{2}-d+1(\mbox{mod }12).\]

\begin{enumerate}
\item $d\equiv0(\mbox{mod }12)$: $w\equiv1(\mbox{mod }12)$, $w-d\equiv1(\mbox{mod }12).$
\item $d\equiv1(\mbox{mod }12)$: $w\equiv1(\mbox{mod }12)$, $w-d\equiv0(\mbox{mod }12).$
\item $d\equiv3(\mbox{mod }12)$: $w\equiv7(\mbox{mod }12)$, $w-d\equiv4(\mbox{mod }12).$
\item $d\equiv4(\mbox{mod }12)$: $w\equiv1(\mbox{mod }12)$, $w-d\equiv9(\mbox{mod }12).$
\item $d\equiv6(\mbox{mod }12)$: $w\equiv7(\mbox{mod }12)$, $w-d\equiv1(\mbox{mod }12).$
\item $d\equiv7(\mbox{mod }12)$: $w\equiv7(\mbox{mod }12)$, $w-d\equiv0(\mbox{mod }12).$
\item $d\equiv9(\mbox{mod }12)$: $w\equiv1(\mbox{mod }12)$, $w-d\equiv4(\mbox{mod }12).$
\item $d\equiv10(\mbox{mod }12)$: $w\equiv7(\mbox{mod }12)$, $w-d\equiv9(\mbox{mod }12).$
\end{enumerate}
This proves the first part of the theorem.

Now we will find the divisibility conditions of the second and the
third parts of the theorem.

Note that by using the expression for $a$:

\[
a=\frac{w-d^{2}+d-1}{d-1},\]
we can represent the size of the sphere as follows:\[
\Phi_{1}(w,a)=1+w(w+a)=(w+a+d)(w-d+1).\]
The code $C$ is a $t-(n,w,\lambda_{t})$-design for each $t,\;0\leq t\leq w-d=f(w,a)$,
where \[
\lambda_{t}=\frac{\binom{n-t}{w-t}}{\Phi_{1}(n,w)}=\frac{\binom{2w+a-t}{w+a}}{(w+a+d)(w-d+1)}.\]
Let denote \[
\lambda:=\lambda_{w-d}=\frac{\binom{w+a+d}{w+a}}{(w+a+d)(w-d+1)}.\]
We simplify the expression for $\lambda$, by using that $w+a+d-1=\frac{w(d-1)+w-d^{2}+d-1+(d-1)^{2}}{d-1}=\frac{wd-d}{d-1}$:

\begin{align*}
\lambda & =\frac{\binom{w+a+d-1}{d-1}}{d(w-d+1)}=\frac{\binom{\frac{wd-d}{d-1}}{d-1}}{d(w-d+1)}=\frac{\left(\frac{wd-d}{d-1}\right)!}{(d-1)!\left(\frac{wd-d}{d-1}-(d-1)\right)!d(w-d+1)}\\
 & =\frac{\left(\frac{wd-d}{d-1}\right)\left(\frac{wd-d}{d-1}-1\right)...\left(\frac{wd-d}{d-1}-(d-2)\right)}{(d-1)!d(w-d+1)}\\
 & =\frac{(wd-d)(wd-d-(d-1))...(wd-d-(d-2)(d-1))}{(d-1)!(d-1)^{d-1}d(w-d+1)}\end{align*}
Thus we get the first divisibility condition:\[
\lambda=\frac{\prod_{i=0}^{d-2}(wd-(d+i(d-1)))}{(d-1)!(d-1)^{d-1}d(w-d+1)}\in\mathbb{Z}.\]
The code $C$ is a $t-(n,w,\lambda_{t})$-design for each $t,\;0\leq t\leq w-d=f(w,a)$,
therefore for all $t$, $0\leq t\leq w-d$

\[
\Phi_{1}(w,a)\left|\binom{2w+a-t}{w-t}\right.\]
or for all $0\leq s\leq w-d$, \[
\Phi_{1}(w,a)\left|\binom{w+a+d+s}{w+a}\right..\]
Note that

\[
\binom{w+a+d+s}{w+a}=\binom{w+a+d}{w+a}\frac{(w+a+d+1)(w+a+d+2)...(w+a+d+s)}{(d+1)(d+2)...(d+s)},\]
where $0\leq s\leq w-d$.

Note also that \[
\frac{\binom{w+a+d}{w+a}}{\Phi_{1}(w,a)}=\lambda,\]
therefore, the last condition can be rewritten as follows:

\[
\lambda\frac{(w+a+d+1)(w+a+d+2)...(w+a+d+s)}{(d+1)(d+2)...(d+s)}\in\mathbb{Z},\]
for all $0\leq s\leq w-d$.

Since $w+a+d+s=\frac{wd+sd-(s+1)}{d-1}$ we finally get the second
divisibility condition:

\[
\lambda\prod_{j=1}^{s}\left[\frac{wd+jd-(j+1)}{(d-1)(d+j)}\right]\in\mathbb{Z},\]
for all $0\leq s\leq w-d$ , where \[
\lambda=\frac{\prod_{i=0}^{d-2}(wd-(d+i(d-1)))}{(d-1)!(d-1)^{d-1}d(w-d+1)}\in\mathbb{Z}\]

\hfill{}$\square$

\subsection{Improvement of Roos' bound for $1$-perfect codes}

From the Roos' bound , it follows that if a $1$-perfect code exists
in $J(2w+a,w)$, then \[
2w+a\leq3(w-1)\]
or \[
a\leq w-3.\]

Now we use the divisibility conditions from the previous section in
order to improve this bound.

\noindent \textbf{Theorem 40.} If a $1$-perfect code exists in $J(2w+a,w)$,
then \[
a<\frac{w}{11}.\]

\noindent \emph{Proof.} Assume that there exists a $1$-perfect code
$C$ in $J(2w+a,w)$ and that the strength of $C$ is $w-d$. Then,
by Theorem 38 \begin{equation}
\lambda=\frac{\prod_{i=0}^{d-2}(wd-(d+i(d-1)))}{(d-1)!(d-1)^{d-1}d(w-d+1)}\in\mathbb{Z}.\label{eq:2}\end{equation}
Given $w=6k+1$ for some integer $k$, we rewrite the expression for
$\lambda$ as follows:

\begin{equation}
\lambda=\frac{\prod_{i=0}^{d-2}(6kd-i(d-1))}{(d-1)!(d-1)^{d-1}(d(6k+1)-d^{2}+d)}.\label{eq:3}\end{equation}
Since $d-1$$\left|6k\right.$, we rewrite the last expression as\[
\lambda=\frac{\left(d\frac{6k}{d-1}\right)\left(d\frac{6k}{d-1}-1\right)...\left(d\frac{6k}{d-1}-(d-2)\right)}{(d-1)!(d(6k+1)-d^{2}+d)}.\]
Note, that the numerator contains $d-1$ successive numbers, therefore
$(d-1)!$ divides it. In addition, $d-1$ does not divide $d(6k+1)-d^{2}+d$,
because $\textrm{\mbox{ }gcd}(d-1,d)=1$ and $\mbox{gcd}(d-1,6k+1)=1,$
therefore we should determine if $d(6k+1)-d^{2}+d$ divides the numerator
of (\ref{eq:3}), or if $d(w-d+1)$ divides the numerator of (\ref{eq:2}).
Note also that the size of the sphere must be squarefree \cite{Gordon},
in particular the expression $w-d+1$ must be squarefree as a factor
of $\Phi_{1}$.

Now we examine several first values of $d>1$.

\begin{itemize}
\item $d=3$. From (\ref{eq:2}) \[
\lambda=\frac{(3w-3)(3w-5)}{2!2^{2}3(w-2)}=\frac{(w-1)(3w-5)}{8(w-2)},\]

\end{itemize}
and since gcd$(w-1,w-2)=1$ and gcd$(3w-5,3w-6)=1,$ $\lambda\notin\mathbb{Z}$.
Contradiction.

Therefore, $d>3$ and $a\leq\frac{w-4^{2}+4-1}{3}<\frac{w}{3}$.

\begin{itemize}
\item $d=4.$ From (\ref{eq:1})\[
\lambda=\frac{4(w-1)(4w-7)(4w-10)}{3!3^{3}4(w-3)},\]

\end{itemize}
therefore, all possible factors of $w-3$ are $2$ and $5$, but $a=\frac{w-13}{3}$,
thus $w>13$. Contradiction.

Therefore, $d>4$ and $a\leq\frac{w-6^{2}+6-1}{5}<\frac{w}{5}$.

\begin{itemize}
\item $d=6$. From (\ref{eq:2})\[
\lambda=\frac{6(w-1)(6w-11)(6w-16)(6w-21)(6w-26)}{5!5^{5}6(w-5)},\]

\end{itemize}
therefore, all possible factors of $w-5$ are $2$, $19$, $7$ and
$3$. But $w\equiv1(\mbox{mod }6),$ hence $w-5\equiv2(\mbox{mod }6),$so
$w-5=2*7$, or $w-5=2*19$, or $w-5=2*7*19,$ therefore $w=19$, 43
or 271. But $a=\frac{w-31}{5},$ so the only possible value for $w$
is 271 and $a=48$. But it must be that $\Phi_{1}(n,w)\left|\binom{n-i}{w-i}\right.$for
all $0\leq i\leq w-6$, and for $i=w-7$ it is false.

Therefore, $d>6$, and $a\leq\frac{w-7^{2}+7-1}{6}<\frac{w}{6}$.

\begin{itemize}
\item $d=7$. From (\ref{eq:2})\[
\lambda=\frac{7(w-1)(7w-13)(7w-19)(7w-25)(7w-31)(7w-37)}{6!*6^{6}*7(w-6)},\]

\end{itemize}
therefore, all possible factors of $w-6$ are $5,29,23,17$ or $11$.
Since $w-6\equiv1(\mbox{mod }6)$ and all possible factors are $-1(\mbox{mod }6)$,
the number of factors of $w-6$ is even. Note that $w-6\equiv1(\mbox{mod }4)$,
all factors are $\pm1(\mbox{mod }4)$, and only $23\equiv-1(\mbox{mod}4)$
and $11\equiv-1(\mbox{mod}4)$. Thus $23$ and $11$ either appear
together or do not appear at all.

Given that $w=6k+1$, $k$ is integer, then\[
\left.12\right|a=\frac{6k-42}{6}=k-7,\]
therefore, \begin{align}
k\equiv7(\mbox{mod }12).\label{eq:17}\end{align}
Thus all possible cases are:

\begin{enumerate}
\item 4 factors: $w-6=6k-5=$$23*11*5*29$, $23*11*5*17$ or $23*11*29*17$.
In any case we obtain contradiction to (\ref{eq:17}), except for
$w-6=$$23*11*5*29$, in which case $w=36691,$ $a=6108.$ Here $\Phi_{1}(n,w)$
does not divide $\binom{n-i}{w-i}$ for $i=w-11$.
\item 2 factors: $w-6=6k-5=23*11$, $5*29,$ $5*17$ or $29*17$. In any
case we obtain contradiction to (\ref{eq:17}), except for $w-6=23*11$,
in which case $w=259,$ $a=36.$ Here $\Phi_{1}(n,w)$ does not divide
$\binom{n-i}{w-i}$ for $i=w-8$.
\end{enumerate}
In any case we obtain contradiction, therefore, $d>7$ and $a\leq\frac{w-9^{2}+9-1}{8}<\frac{w}{8}$.

\begin{itemize}
\item $d=9$. From (\ref{eq:2})\[
\lambda=\frac{9(w-1)(9w-17)(9w-25)(9w-33)(9w-41)(9w-49)(9w-57)(9w-65)}{8!*8^{8}*9(w-8)},\]

\end{itemize}
therefore, all possible factors of $w-8$ are $5,7,11,13,23,31$ and
$47$.

Note that $\left.12\right|a=\frac{6k-72}{8}$, so $\left.12*8\right|6k-72$,
therefore, $k=16b+12$, for some integer $b$, $w=96b+73$, $w-8\equiv$$1(\mbox{mod }16)\equiv5(\mbox{mod }12).$

Note that $5\equiv5(\mbox{mod }12)$, $7\equiv23\equiv-5(\mbox{mod }12)$,
$11\equiv23\equiv47\equiv-1(\mbox{mod }12),$ $13\equiv1(\mbox{mod }12)$,
$23\equiv7(\mbox{mod }16)$, $31\equiv47\equiv-1(\mbox{mod }16)$,$11\equiv-5(\mbox{mod }16)$.

Thus all possible cases are:

\begin{enumerate}
\item 6 factors: $w-8=5*7*13*23*31*47$, in this case $\Phi_{1}(n,w)$ does
not divide $\binom{n-i}{w-i}$ for $i=w-10$. Contradiction.
\item 5 factors: $w-8=5*7*11*31*47$ or $7*11*13*23*47$. In both cases
$\Phi_{1}(n,w)$ does not divide $\binom{n-i}{w-i}$ for $i=w-10$.
Contradiction.
\item 3 factors: $w-8=5*11*23$ or $11*13*31$. In both cases $\Phi_{1}(n,w)$
does not divide $\binom{n-i}{w-i}$ for $i=w-11$. Contradiction.
\item 2 factors: $w-8=5*13$, $7*23$ or $31*47$, In the first case $\Phi_{1}(n,w)$
does not divide $\binom{n-i}{w-i}$ for $i=w-13$, and in the last
two cases $\Phi_{1}(n,w)$ does not divide $\binom{n-i}{w-i}$ for
$i=w-10$. Contradiction.
\end{enumerate}
Therefore, $d>9$ and $a\leq\frac{w-10^{2}+10-1}{9}<\frac{w}{9}$.

\begin{itemize}
\item $d=10$. From (\ref{eq:2})\begin{align*}
\lambda & =\frac{10}{9!*9^{9}*10(w-9)}[(w-1)(10w-19)(10w-28)(10w-37)(10w-46)\\
 & *(10w-55)(10w-64)(10w-73)(10w-82)],\end{align*}

\end{itemize}
therefore, all possible factors of $w-9$ are $2,71,31,53,11,7,13$
and $17$.

Note that $\left.12\right|a=\frac{w-91}{9}=\frac{12k+7-91}{9}=\frac{12k-84}{9},$
thus $\left.12*9\right|12k-84,$ or $\left.9\right|k-7,$ so we can
write $k=9b+7$, for some integer $b$. Also $w-9\equiv1(\mbox{mod }9)\equiv2(\mbox{mod }4)\equiv-2(\mbox{mod }6)\equiv-2(\mbox{mod }12).$
If we consider all possible factors modulo $9,4,6$ and $12$, we
get several constraints, therefore the only possible cases are:

\begin{enumerate}
\item 5 factors: $w-9=2*7*13*17*31$ , $2*17*31*53*71$, $2*7*13*31*53$
or $2*7*11*17*53$. In the two first cases 12 does not divides $a$.
In the third case $\Phi_{1}(n,w)$ does not divide $\binom{n-i}{w-i}$
for $i=w-14$. In the last case $\Phi_{1}(n,w)$ does not divide $\binom{n-i}{w-i}$
for $i=w-11$. Contradiction.
\item 3 factors: $w-9=2*7*11,$ $2*13*17$, $2*13*53$, or $2*31*71$. In
the first two cases 12 does not divide $a$. In the last two cases
$\Phi_{1}(n,w)$ does not divide $\binom{n-i}{w-i}$ for $i=w-13$.
Contradiction.
\end{enumerate}
Therefore, $d>10$. Moreover, since $d\equiv0,1(\mbox{mod }3)$, $d\geq12$.

Conclusion: \[
a\leq\frac{w-12^{2}+12-1}{11}=\frac{w-133}{11}<\frac{w}{11}.\]

\hfill{}$\square$

Note, that while we do not show a generalization, we can further improve
the bound on $a$ by applying this technique.

\subsection{Number theory's constraints for size of $\Phi_{1}(n,w)$}

In \cite{conf_dist} Etzion shown that if $1$-perfect $C$ code exists
in $J(n,w)$, then\[
w=(\beta-\alpha)(\beta+\alpha+1)+1,\]
\[
n=2(\beta-\alpha)(\beta+\alpha+1)+2\alpha+2,\]
and the strength of the code $C$ is\[
(\beta-\alpha)(\beta+\alpha),\]
where $2\alpha=n-2w$ and $2\beta+1=\sqrt{(n-2w+1)^{2}+(w-1)}$.

\noindent \textbf{Lemma 41.} If $1$-perfect code exists in $J(2w+a,w)$
then\[
\Phi_{1}(w,a)=(\beta^{2}-\alpha^{2}+1)((\beta+1)^{2}-\alpha^{2}+1),\]

\begin{itemize}
\item gcd$\left((\beta^{2}-\alpha^{2}+1)((\beta+1)^{2}-\alpha^{2}+1)\right)=1$
\item $\beta^{2}-\alpha^{2}+1$ is squarefree
\item $(\beta+1)^{2}-\alpha^{2}+1)$ is squarefree
\end{itemize}
where $2\alpha=n-2w$ and $2\beta+1=\sqrt{(n-2w+1)^{2}+(w-1)}$.

\noindent \emph{Proof}\textbf{.} In the proof of the Theorem 38 it
was shown that if $1$-perfect code exists in $J(2w+a,w)$, and its
strength is $w-d$ for some integer $d$, then \[
\Phi_{1}(w,a)=(w-d+1)(w+a+d).\]
Since $d=w-(\beta-\alpha)(\beta+\alpha)=(\beta-\alpha)(\beta+\alpha+1)+1-(\beta-\alpha)(\beta+\alpha)=\beta-\alpha+1$,

$w-d+1=(\beta-\alpha)(\beta+\alpha)+1=\beta^{2}-\alpha^{2}+1$,

$w+a+d=(\beta-\alpha)(\beta+\alpha+1)+1+2\alpha+\beta-\alpha+1=(\beta+1)^{2}-\alpha^{2}+1$,

\noindent the expression for $\Phi_{1}(w,a)$ is\[
\Phi_{1}(w,a)=(w-d+1)(w+a+d).\]
Gordon \cite{Gordon} proved, that $\Phi_{1}(w,a)$ must be squarefree,
which proves the lemma.

\hfill{}$\square$

\section{Moments}

\subsection{Introduction}

\subsubsection{Configuration distribution}

The following definitions appear in \cite{conf_dist}.

Let $C$ be a code in $J(n,w)$. We can partition the coordinate set
$N$ into $r$ subsets $\left\{ \alpha_{1},\alpha_{2},...,\alpha_{r}\right\} .$
A vector $x\in V_{w}^{n}$ can be written as $x=(x_{1},x_{2},...,x_{r})$,
where $x_{i}\in\alpha_{i}$, $1\leq i\leq r$. We say that $x$ is
from \emph{configuration} $(w_{1},w_{2},...,w_{r})$, $\sum_{i=1}^{r}w_{i}=w$,
if $\left|x_{i}\right|=w_{i}$, $1\leq i\leq r$. We denote by $D_{(w_{1},w_{2},...,w_{r})}$
the number of codewords from configuration $(w_{1},w_{2},...,w_{r})$.
The \emph{configuration distribution of $C$} is a vector consisting
of all the values $D_{(w_{1},w_{2},...,w_{r})}$, where $w_{i}\leq\left|\alpha_{i}\right|$,
$1\leq i\leq r$, and $\sum_{i=1}^{r}w_{i}=w$.

In \cite{nonexist} several partitions with $r=2$ were considered.
The most important one is the one in which $\left|\alpha_{1}\right|=w$
and $\left|\alpha_{2}\right|=w+a$. Clearly, permutation on the columns
of $e$-perfect code $C$ will result in an $e$-perfect code isomorphic
to $C$. In this case it was proved in \cite{nonexist} that an $e$
-perfect code have exactly $e+1$ different configuration distributions.

In order to avoid confusion we will assume that the vector from configuration
$(w,0)$ is always a codeword in a perfect code $C$. If we permute
the columns of $C$ (in other words, we take another partition $\left\{ \beta_{1},\beta_{2}\right\} $
of $N$, such that $\left|\beta_{1}\right|=w$ and $\left|\beta_{2}\right|=w+a$)
in a way that the vector from configuration $(w,0)$ is not a codeword
we will call the obtained code a \emph{translate} of $C$. For each
$j$, $1\leq j\leq e$, there exists a translate with exactly one
\emph{translate-word} from configuration $(w-j,j)$, and no translate
-word from configuration $(w-i,i)$, $0\leq i\leq e$, $i\neq j$.
The translate -word from configuration $(w-j,j)$ will be called a
\emph{translate leader}.

Let $A_{i}$, $0\leq i\leq w$, be the number of codewords in configuration
$(w-i,i)$ and let $B_{i,j}$, $0\leq i\leq w$, $0\leq j\leq e$,
be the number of translate-words from configuration $(w-i,i)$ in
the translate with translate-leader $(w-j,j)$. Note, that $A_{i}=B_{i,0}$
and $B_{i,j}=D_{(w-i,i)}$in the corresponding translate. $A_{i}$
is also the number of codewords which have distance $i$ to the codeword
from configuration $(w,0)$ and $(A_{i})_{i=0}^{w}$ is the inner
distance distribution of the code in the Johnson scheme. $(B_{i})_{i=0}^{w}$
is the configuration distribution which is akin to the weight distribution
in the Hamming scheme.

Etzion in \cite{conf_dist} proved the following theorem:

\noindent \textbf{Theorem 42.} For a given $e$-perfect code $C$
in $J(n,w)$ we have\[
\sum_{j=0}^{e}\binom{w}{j}\binom{w+a}{j}B_{i,j}=\binom{w}{i}\binom{w+a}{i}.\]

\subsubsection{Moments.}

In \cite{conf_dist} Etzion defined a generalization for \emph{moments}
of a code which was given for the Hamming scheme \cite{moments}.

Let $C$ be an $e$-perfect code in $J(n,w)$, and let $\left\{ \alpha_{1},\alpha_{2}\right\} $
be a partition of $N$ such that $\left|\alpha_{1}\right|=k$ and
$\left|\alpha_{2}\right|=n-k$. Let $A_{i}$ be the number of codewords
from configuration $(i,w-i)$ (note, that this definition is slightly
different from the one in the previous definition). Let $\left\{ \beta_{1},\beta_{2}\right\} $
be another partition of $N$ such that $\left|\beta_{1}\right|=k$
and $\left|\beta_{2}\right|=n-k$, and $B_{i}$ be the number of codewords
from configuration $(i,w-i)$ with respect to this partition.

The $r$-th \emph{power moment}, \emph{$0\leq r$,} of $C$ with respect
to these partitions is defined by\[
\sum_{i=0}^{k}i^{r}A_{i},\:\sum_{i=0}^{k}i^{r}B_{i}\]
and the $r$-th \emph{binomial moment}, \emph{$0\leq r$,} of $C$
is defined by\[
\sum_{i=0}^{k}\binom{i}{r}A_{i},\:\sum_{i=0}^{k}\binom{i}{r}B_{i}.\]

We define the \emph{difference configuration distributions} between
the two partitions by $\Delta_{i}=A_{i}-B_{i}$, $0\leq i\leq k$.
The $r$-th power moments and the $r$-th binomial moments with respect
to the difference configuration distributions are defined by\[
\sum_{i=0}^{k}i^{r}\Delta_{i},\:\sum_{i=0}^{k}\binom{i}{r}\Delta_{i}.\]

Two types of moments are connected by \emph{Stirling number of the
second kind} $S(r,v)$. $S(r,v)$, $r\geq v\geq0$ is the number of
ways to partition a set of $r$ elements into $v$ nonempty sets.
The following are known three formulas \cite{stirling} :

\[
S(r,v)=\frac{1}{v!}\sum_{i=0}^{r}(-1)^{v-i}\binom{v}{i}\: i^{r},\]

\[
S(r,v)=S(r-1,v-1)+vS(r-1,v),\]
where $S(r,1)=S(r,r)=1$ and $S(r,0)=0$ for $r>0$,

\[
i^{r}=\sum_{v=0}^{r}v!\binom{i}{v}S(r,v).\]
Hence \[
\sum_{i=0}^{k}i^{r}\Delta_{i}=\sum_{i=0}^{k}\sum_{v=0}^{r}v!\binom{i}{v}S(r,v)\Delta_{i}=\sum_{v=0}^{r}v!S(r,v)\sum_{i=0}^{k}\binom{i}{v}\Delta_{i}.\]
Therefore, it can be proved by induction that

\noindent \textbf{Theorem} \textbf{43}. For a given integer $t$,
$\sum_{i=0}^{k}i^{r}\Delta_{i}=0$ for all $0\leq r\leq t$ if and
only if $\sum_{i=0}^{k}\binom{i}{r}\Delta_{i}=0$ for all $0\leq r\leq t$.

In \cite{conf_dist} Etzion showed that for $r\leq\varphi$, where
$\varphi$ is a strength of the code, the values of the binomial moments
can be easily computed.

\noindent \textbf{Lemma 44.} If $C$ is a perfect code in $J(n,w)$
and $\varphi$ is its strength, then for each $r$, $0\leq r\leq\varphi$
we have\[
\sum_{i=0}^{k}\binom{i}{r}A_{i}=\sum_{i=0}^{k}\binom{i}{r}B_{i}=\binom{k}{r}\frac{\binom{n-r}{w-r}}{\Phi_{e}(n,w)}.\]

\noindent \textbf{Corollary 45.} If $C$ is a perfect code in $J(n,w)$
and $\varphi$ is its strength, then for each $r$, $0\leq r\leq\varphi$
we have $\sum_{i=0}^{k}\binom{i}{r}\Delta_{i}=0$ and $\sum_{i=0}^{k}i^{r}\Delta_{i}=0$.

\subsection{Binomial moments for $1$-perfect codes in $J(n,w)$ }

We saw in the previous section that for $r\leq\varphi$, where $\varphi$
is a strength of the code, the values of the binomial moments can
be easily computed. In this section we consider the binomial moments
for $r>\varphi$, for $1$-perfect codes in $J(n,w).$

In \cite{conf_dist} Etzion proved the following lemma.

\noindent \textbf{Lemma 46.} Given $\left\{ H_{1},H_{2}\right\} $
partition of $N$ such that $|H_{1}|=k$, $|H_{2}|=n-k$, for any
$i$, $0\leq i\leq k$ we have\[
(i+1)(w+a-k+i+1)A_{i+1}+[(1+i(k-1)+(w-i)(w+a-k+i)]A_{i}\]

\[
+(k-i+1)(w-i+1)A_{i-1}=\binom{k}{i}\binom{2w+a-k}{w-i},\]
where $A_{i}$ is the number of codewords from configuration $(i,w-i)$.

Let $\left\{ \alpha_{1},\alpha_{2}\right\} $ be a partition of $N$
such that $|\alpha_{1}|=w$, $|\alpha_{2}|=n-w$, and a vector of
$(w,\:0)$ configuration be a codeword. Let $A_{i}$ be the number
of codewords from configuration $(i,w-i)$. Let $\left\{ \beta_{1},\beta_{2}\right\} $
be another partition of $N$ such that $|\beta_{1}|=w$, $|\beta_{2}|=n-w$,
let $B_{i}$ be the number of codewords from configuration $(i,w-i)$
with respect to this partition, and let $\Delta_{i}=A_{i}-B_{i}$,
$0\leq i\leq w$.

\noindent \textbf{Theorem 47.} If $C$ is a $1$-perfect code in $J(n,w)$
and $\varphi$ is its strength, then for each $k$, $\varphi<k\leq w$,
we have\[
\sum_{i=0}^{w}\binom{i}{k}\Delta_{i}=(-1)^{w-k}\prod_{l=1}^{w-k}\frac{[(l-1)n+l^{2}-l+1-w(2l-1)]}{l^{2}}\]

\[
\sum_{i=0}^{w}\binom{i}{k}B_{i}=\frac{\binom{n-w}{k}\binom{n-k}{w-k}-(-1)^{w-k}\prod_{l=1}^{w-k}\frac{[(l-1)n+l^{2}-l+1-w(2l-1)]}{l^{2}}}{\Phi_{1}(n,w)}\]

\[
\sum_{i=0}^{w}\binom{i}{k}A_{i}=\frac{w(n-w)(-1)^{w-k}\prod_{l=1}^{w-k}\frac{[(l-1)n+l^{2}-l+1-w(2l-1)]}{l^{2}}+\binom{n-w}{k}\binom{n-k}{w-k}}{\Phi_{1}(n,w)}.\]

\noindent \emph{Proof}\textbf{.} Assume that $C$ is a 1-perfect code
in $J(n,w)$ and $\varphi$ is its strength. By Lemma 46 we have\begin{eqnarray*}
\binom{w}{i}\binom{n-w}{w-i} & = & A_{i+1}(i+1)(i+1+n-2w)\\
 & + & A_{i}(1+(w-i)(n-2w+2i))+A_{i-1}(w-i+1)^{2}\end{eqnarray*}

\begin{eqnarray*}
\binom{w}{i}\binom{n-w}{w-i} & = & B_{i+1}(i+1)(i+1+n-2w)\\
 & + & B_{i}(1+(w-i)(n-2w+2i))+B_{i-1}(w-i+1)^{2}\end{eqnarray*}
where $0\leq i\leq w$.

Therefore,\begin{eqnarray*}
0 & = & \Delta_{i+1}(i+1)(i+1+n-2w)+\Delta_{i}(1+(w-i)(n-2w+2i))\\
 & + & \Delta_{i-1}(w-i+1)^{2},\end{eqnarray*}
or

\begin{eqnarray*}
0 & = & \Delta_{i+1}[(i+1)^{2}+(n-2w)(i+1)]+\Delta_{i}[1+w(n-2w)+i(4w-n)-2i^{2}]\\
 & + & \Delta_{i-1}[w^{2}-2w(i-1)+(i-1)^{2}]\end{eqnarray*}
Multiply it by $\binom{i}{k}$ and sum over all $i$, $0\leq i\leq w$
:

\begin{eqnarray*}
0 & = & \sum_{i=0}^{w}\binom{i}{k}(i+1)^{2}\Delta_{i+1}+(n-2w)\sum_{i=0}^{w}\binom{i}{k}(i+1)\Delta_{i+1}+\sum_{i=0}^{w}\binom{i}{k}\Delta_{i}\\
 & + & w(n-2w)\sum_{i=0}^{w}\binom{i}{k}\Delta_{i}+(4w-n)\sum_{i=0}^{w}\binom{i}{k}\, i\Delta_{i}\\
 & - & 2\sum_{i=0}^{w}\binom{i}{k}\, i^{2}\Delta_{i}+w^{2}\sum_{i=0}^{w}\binom{i}{k}\Delta_{i-1}\\
 & - & 2w\sum_{i=0}^{w}\binom{i}{k}(i-1)\Delta_{i-1}+\sum_{i=0}^{w}\binom{i}{k}(i-1)^{2}\Delta_{i-1}\\
\\\end{eqnarray*}
We prove the following proposition (see Appendix A).

\noindent \textbf{Proposition 48.} For each $k$, $\varphi<k\leq w$,
we have \begin{equation}
0=[1+k^{2}-k(1+n)+nw-w^{2}]\sum_{i=0}^{w}\binom{i}{k}\Delta_{i}+(1-k+w)^{2}\sum_{i=0}^{w}\binom{i}{k-1}\Delta_{i}.\label{eq:4}\end{equation}
Note that $\sum_{i=0}^{w}\binom{i}{w}\Delta_{i}=\Delta_{w}=1$.

\noindent If we assume that $k=w$, from (\ref{eq:4}) we get:

\begin{eqnarray*}
0 & = & (1-w)\sum_{i=0}^{w}\binom{i}{w}\Delta_{i}+\sum_{i=0}^{w}\binom{i}{w-1}\Delta_{i}\end{eqnarray*}
therefore,\begin{eqnarray*}
\sum_{i=0}^{w}\binom{i}{w-1}\Delta_{i} & = & -(1-w)\end{eqnarray*}
If we assume that $k=w-1$, from (\ref{eq:4}) we get:\begin{eqnarray*}
0 & = & (3+n-3w)\sum\binom{i}{w-1}\Delta_{i}+2^{2}\sum\binom{i}{w-2}\Delta_{i}\end{eqnarray*}
therefore, \begin{eqnarray*}
\sum_{i=0}^{w}\binom{i}{w-2}\Delta_{i} & = & \frac{(n+3-3w)(1-w)}{2^{2}}\end{eqnarray*}
In general, for $k=w-j$ from (\ref{eq:4}) we get:\begin{eqnarray*}
0 & = & [jn+(j^{2}+j+1)-w(2j+1)]\sum_{i=0}^{w}\binom{i}{w-j}\Delta_{i}+(j+1)^{2}\sum_{i=0}^{w}\binom{i}{w-j-1}\Delta_{i}\end{eqnarray*}
Therefore,\begin{eqnarray*}
\sum_{i=0}^{w}\binom{i}{w-j-1}\Delta_{i} & =- & \frac{[jn+(j^{2}+j+1)-w(2j+1)]\sum_{i=0}^{w}\binom{i}{w-j}\Delta_{i}}{(j+1)^{2}}\\
\sum_{i=0}^{w}\binom{i}{w-j}\Delta_{i} & = & (-1)^{j}\prod_{l=1}^{j}\frac{[(l-1)n+l^{2}-l+1-w(2l-1)]}{l^{2}}\end{eqnarray*}
or\[
\sum_{i=0}^{w}\binom{i}{k}\Delta_{i}=(-1)^{w-k}\prod_{l=1}^{w-k}\frac{[(l-1)n+l^{2}-l+1-w(2l-1)]}{l^{2}}\]
 for $k=\varphi+1,...,w$, where $\varphi$ is the strength of the
code.

Since $\Delta_{i}=A_{i}-B_{i}$ and by Theorem 42 $\binom{w}{i}\binom{n-w}{i}=A_{i}+w(n-w)B_{i}$,
we have:

\begin{eqnarray*}
\sum_{i=0}^{w}\binom{i}{k}\Delta_{i} & = & \sum_{i=0}^{w}\binom{i}{k}A_{i}-\sum_{i=0}^{w}\binom{i}{k}B_{i}\\
 & = & \sum_{i=0}^{w}\binom{i}{k}\binom{w}{i}\binom{n-w}{i}-w(n-w)\sum_{i=0}^{w}\binom{i}{k}B_{i}-\sum_{i=0}^{w}\binom{i}{k}B_{i}\end{eqnarray*}
therefore, since $\sum_{i=0}^{w}\binom{i}{k}\binom{w}{i}\binom{n-w}{i}=\binom{n-w}{k}\binom{n-k}{w-k}$,

\begin{eqnarray*}
\sum_{i=0}^{w}\binom{i}{k}B_{i} & = & \frac{\sum_{i=0}^{w}\binom{i}{k}\binom{w}{i}\binom{n-w}{i}-\sum_{i=0}^{w}\binom{i}{k}\Delta_{i}}{w(n-w)+1}=\frac{\binom{n-w}{k}\binom{n-k}{w-k}-\sum_{i=0}^{w}\binom{i}{k}\Delta_{i}}{w(n-w)+1}\end{eqnarray*}

\[
=\frac{\binom{n-w}{k}\binom{n-k}{w-k}-(-1)^{w-k}\prod_{l=1}^{w-k}\frac{[(l-1)n+l^{2}-l+1-w(2l-1)]}{l^{2}}}{w(n-w)+1}\]

\[
\sum_{i=0}^{w}\binom{i}{k}A_{i}=\frac{w(n-w)(-1)^{w-k}\prod_{l=1}^{w-k}\frac{[(l-1)n+l^{2}-l+1-w(2l-1)]}{l^{2}}+\binom{n-w}{k}\binom{n-k}{w-k}}{w(n-w)+1}\]
where $k=\varphi+1,...,w$, and $\varphi$ is the strength of the
code.

\hfill{}$\square$

Note that if in expression (\ref{eq:4}) we assume that $k=\varphi+1$,
then the second summand disappears, and the coefficient of the first
summand must be $0$. Therefore, we got equation for $\varphi$, and
its solution gives us the expression for the strength of a $1$-perfect
code:\[
\varphi=\frac{n-1-\sqrt{(n-2w+1)^{2}+4(w-1)}}{2}.\]
Therefore, binomial moments is a second way to get the strength of
the perfect code.

\subsubsection{Applications of Binomial moments for $1$-perfect codes in $J(n,w)$}

Now we consider the $(w-5)$-binomial moment and several partitions
of set of coordinates $N$ in order to exclude a number of parameters
for $1$ -perfect code.

We examine the $(w-5)$-binomial moment with respect to the difference
configuration distributions:

\[
\sum\binom{i}{w-5}\Delta_{i}=\frac{(w-1)(a-w+3)(2a-w+7)(3a-w+13)(4a-w+21)}{(5!)^{2}}\]

Note that binomial moments must be integer number, therefore we have
one of divisibility conditions for $1$-perfect code.

In addition we examine the following three partitions of set of coordinates:

\begin{enumerate}
\item $\left\{ \alpha_{1},\alpha_{2}\right\} $, such that $|\alpha_{1}|=w$,
$|\alpha_{2}|=n-w$, and the vector of $(w,0)$ configuration is a
codeword. Let $A_{i}$ be the number of codewords from configuration
$(i,w-i)$ with respect to this partition. By Lemma 46 and using the
fact that $A_{w}=1$, $A_{w-1}=0$ we obtain the following expression\begin{eqnarray*}
A_{w-5} & = & \frac{w(w-1)(w+a)(w+a-1)}{(5!)^{2}}[a^{2}(26+(w-9)w)\\
 & + & (w-3)(-181+w(87+(w-15)w))\\
 & + & a(-221+w(132+w(2w-27)))]\end{eqnarray*}

\item $\left\{ \beta_{1},\beta_{2}\right\} $, such that $|\beta_{1}|=w-2$,
$|\beta_{2}|=n-w+2$, and the vector of $(w-2,2)$ configuration is
a codeword. Let $B_{i}$ be the number of codewords from configuration
$(i,w-i)$ with respect to this partition. By Lemma 46 and using the
fact that $B_{w-2}=1$, $B_{w-3}=\frac{(w+a)(w+a-1)}{6}$ we obtain
the following expression\begin{eqnarray*}
B_{w-5} & = & \frac{1}{15*48}(w+a-1)(w+a)[a^{2}(26+(w-9)w)\\
 & + & (w-3)(19+w(-3+(w-5)w))+a(-21+w(42+w(2w-17)))]\end{eqnarray*}

\item $\left\{ \gamma_{1},\gamma_{2}\right\} $, such that $|\gamma_{1}|=w+2$,
$|\gamma_{2}|=n-w-2$, and the vector of $(w,0)$ configuration is
a codeword. Let $C_{i}$ be the number of codewords from configuration
$(i,w-i)$ with respect to this partition. By Lemma 46 and using the
fact that $C_{w}=1$, $C_{w-1}=\frac{w(w-1)}{6}$ we obtain the following
expression\begin{eqnarray*}
C_{w-3} & = & \frac{1}{15*48}w(w-1)[a^{2}(-4+(w+1)w)\\
 & + & (w-3)(19+w(-3+(w-5)w+a(49+w(-18+w(2w-7)))))]\end{eqnarray*}

\end{enumerate}
We chose those expressions since one of the factors of all the denumerators
is '5'.

Since we know that $w\equiv w+a\equiv1(\textrm{mod }12)$ or $w\equiv w+a\equiv7(\textrm{mod }12)$
we consider all possible cases for $w$ and $w+a$ modulo 60
.

Using the above four divisibility conditions, we build two tables
$w$ versus $w+a$ modulo 60, where $w\equiv w+a\equiv1(\textrm{mod }12)$
and $w\equiv w+a\equiv7(\textrm{mod }12)$, respectively, where '-'
denotes that there are no $1$-perfect codes with such parameters.

\begin{table}[H]
\caption{\label{tab:table1}$w\equiv w+a\equiv1(\textrm{mod}12)$}

\begin{centering}
{\small }\begin{tabular}{|>{\centering}p{0.15\textwidth}|>{\centering}p{0.05\textwidth}|>{\centering}p{0.05\textwidth}|>{\centering}p{0.05\textwidth}|>{\centering}p{0.05\textwidth}|>{\centering}p{0.05\textwidth}|}
\hline
{\small }\begin{tabular}{>{\raggedright}p{0.01\textwidth}>{\raggedright}p{0.08\textwidth}}
 & {\small w+a}\tabularnewline
{\small w} & \tabularnewline
\end{tabular} & 1 & {\small 13} & {\small 25} & {\small 37} & {\small 49}\tabularnewline
\hline
{\small 1} &  &  & -- &  & {\small --}\tabularnewline
\hline
{\small 13} &  &  & -- & {\small --} & {\small --}\tabularnewline
\hline
{\small 25} &  & {\small --} & -- &  & \tabularnewline
\hline
{\small 37} &  & {\small --} &  & {\small --} & {\small --}\tabularnewline
\hline
{\small 49} & {\small --} & {\small --} &  & {\small --} & {\small --}\tabularnewline
\hline
\end{tabular}\\

\par\end{centering}
\end{table}

\begin{table}[H]
\caption{\label{tab:table2}$w\equiv w+a\equiv7(\textrm{mod}12)$}

\begin{centering}
{\small }\begin{tabular}{|>{\centering}p{0.15\textwidth}|>{\centering}p{0.05\textwidth}|>{\centering}p{0.05\textwidth}|>{\centering}p{0.05\textwidth}|>{\centering}p{0.05\textwidth}|>{\centering}p{0.05\textwidth}|}
\hline
{\small }\begin{tabular}{>{\raggedright}p{0.01\textwidth}>{\raggedright}p{0.08\textwidth}}
 & {\small w+a}\tabularnewline
{\small w} & \tabularnewline
\end{tabular} & 7 & {\small 19} & {\small 31} & {\small 43} & 55\tabularnewline
\hline
7 & -- & -- &  & -- & \tabularnewline
\hline
{\small 19} & -- & -- & -- & {\small --} & \tabularnewline
\hline
31 &  & {\small --} &  &  & \tabularnewline
\hline
43 & -- & {\small --} &  &  & {\small --}\tabularnewline
\hline
55 &  &  &  & {\small --} & {\small --}\tabularnewline
\hline
\end{tabular}\\

\par\end{centering}
\end{table}

In addition, if we write $w=60k+i$ and $w+a=60y+j$ for $i,j\in\{1,13,25,37,49,$
$7,19,31,43,55\}$ then we get the following existence conditions:

\begin{itemize}
\item If there exists $1$-perfect code with $w\equiv w+a\equiv13(\textrm{mod }60)$
then $k+y\equiv3(\textrm{mod }5)$.
\item If there exists $1$-perfect code with $w\equiv25(\textrm{mod }60)$
and $w+a\equiv1(\textrm{mod }60)$ then $y\equiv0(\textrm{mod }5)$.
\item If there exists $1$-perfect code with $w\equiv25(\textrm{mod }60)$
and $w+a\equiv37(\textrm{mod }60)$ then $2k-y\equiv2(\textrm{mod }5)$.
\item If there exists $1$-perfect code with $w\equiv37(\textrm{mod }60)$
and $w+a\equiv25(\textrm{mod }60)$ then $4k-3y\equiv4(\textrm{mod }5)$.
\item If there exists $1$-perfect code with $w\equiv7(\textrm{mod }60)$
and $w+a\equiv55(\textrm{mod }60)$ then $4k-3y\equiv0(\textrm{mod }5)$
and $a\equiv0(\textrm{mod }24)$.
\item If there exists $1$-perfect code with $w\equiv31(\textrm{mod }60)$
and $w+a\equiv55(\textrm{mod }60)$ then $k\equiv2(\textrm{mod }5)$
and $a\equiv0(\textrm{mod }24)$.
\item If there exists $1$-perfect code with $w\equiv43(\textrm{mod }60)$
and $w+a\equiv43(\textrm{mod }60)$ then $k+y\equiv2(\textrm{mod }5)$
and $a\equiv0(\textrm{mod }24)$.
\item If there exists $1$-perfect code with $w\equiv55(\textrm{mod }60)$
and $w+a\equiv7(\textrm{mod }60)$ then $2k-y\equiv0(\textrm{mod }5)$
and $a\equiv0(\textrm{mod }24)$.
\item If there exists $1$-perfect code with $w\equiv55(\textrm{mod }60)$
and $w+a\equiv31(\textrm{mod }60)$ then $y\equiv2(\textrm{mod }5)$
and $a\equiv0(\textrm{mod }24)$.
\end{itemize}

\subsection{Binomial moments for $2$-perfect code in $J(2w,w)$ }

In this section we calculate the expression for $k$-th binomial moments
with respect to the difference configuration distributions for all
$k>\varphi$, where $\varphi$ is a strength of a 2-perfect code in
$J(2w,w)$, and obtain expression for strength of this code.

Let $C$ be a 2-perfect code in $J(2w,w)$. Let $\left\{ \alpha_{1},\alpha_{2}\right\} $
be a partition of $N$ such that $|\alpha_{1}|=w$, $|\alpha_{2}|=w$,
and vector of $(w,0)$ configuration is a codeword. Let $A_{i}$ be
the number of codewords from configuration $(i,w-i)$. Let $\left\{ \beta_{1},\beta_{2}\right\} $
be another partition of $N$ such that $|\beta_{1}|=w$, $|\beta_{2}|=w$,
and let $B_{i}$ be number of codewords from configuration $(i,w-i)$
with respect to this partition. The $k$-th binomial moment, $0\leq k$,
of $C$ is defined by\[
\sum_{i=0}^{k}\binom{i}{k}A_{i},\;\sum_{i=0}^{k}\binom{i}{k}B_{i}.\]
By considering how $\binom{w}{i}\binom{w}{w-i}$ vectors from configuration
$(i,w-i)$ are $2$-covered by $C$ we obtain the following formulas
for any $i$, $0\leq i\leq w$:

\begin{eqnarray*}
\binom{w}{i}^{2} & = & \binom{i+2}{2}^{2}A_{i+2}+\binom{w-i+2}{2}^{2}A_{i-2}\\
 & + & \left[(i+1)^{2}+2(i+1)(w-i-1)\binom{i+1}{2}\right]A_{i+1}\\
 & + & \left[(w-i+1)^{2}+2(i-1)(w-i+1)\binom{w-1+1}{2}\right]A_{i-1}\\
 & + & \left[1+2i(w-i)+2\binom{i}{2}\binom{w-i}{2}+i^{2}(w-i)^{2}\right]A_{i}\end{eqnarray*}

\begin{eqnarray*}
\binom{w}{i}^{2} & = & \binom{i+2}{2}^{2}B_{i+2}+\binom{w-i+2}{2}^{2}B_{i-2}\\
 & + & \left[(i+1)^{2}+2(i+1)(w-i-1)\binom{i+1}{2}\right]B_{i+1}\\
 & + & \left[(w-i+1)^{2}+2(i-1)(w-i+1)\binom{w-1+1}{2}\right]B_{i-1}\\
 & + & \left[1+2i(w-i)+2\binom{i}{2}\binom{w-i}{2}+i^{2}(w-i)^{2}\right]B_{i}\end{eqnarray*}
Let $\Delta_{i}=A_{i}-B_{i}$, for $0\leq i\leq w$. Hence we obtain:

\begin{eqnarray*}
0 & = & \binom{i+2}{2}^{2}\Delta_{i+2}+\binom{w-i+2}{2}^{2}\Delta_{i-2}\\
 & + & \left[(i+1)^{2}+2(i+1)(w-i-1)\binom{i+1}{2}\right]\Delta_{i+1}\\
 & + & \left[(w-i+1)^{2}+2(i-1)(w-i+1)\binom{w-1+1}{2}\right]\Delta_{i-1}\\
 & + & \left[1+2i(w-i)+2\binom{i}{2}\binom{w-i}{2}+i^{2}(w-i)^{2}\right]\Delta_{i}\end{eqnarray*}
Next we multiply it by $\binom{i}{k}$ and sum over all $0\leq i\leq w$:

\begin{eqnarray*}
0 & = & \sum_{i=0}^{w}\binom{i}{k}\binom{i+2}{2}^{2}\Delta_{i+2}+\sum_{i=0}^{w}\binom{i}{k}\binom{w-i+2}{2}^{2}\Delta_{i-2}\\
 & + & \sum_{i=0}^{w}\binom{i}{k}\left[(i+1)^{2}+2(i+1)(w-i-1)\binom{i+1}{2}\right]\Delta_{i+1}\\
 & + & \sum_{i=0}^{w}\binom{i}{k}\left[(w-i+1)^{2}+2(i-1)(w-i+1)\binom{w-i+1}{2}\right]\Delta_{i-1}\\
 & + & \sum_{i=0}^{w}\binom{i}{k}\left[1+2i(w-i)+2\binom{i}{2}\binom{w-i}{2}+i^{2}(w-i)^{2}\right]\Delta_{i}\end{eqnarray*}
We prove the following proposition (see Appendix B).

\noindent \textbf{Proposition 49.} For each $k$, $\varphi<k\leq w$,
we have\begin{eqnarray*}
0 & = & \frac{1}{4}(4+k^{4}+5w^{2}-2w^{3}+w^{4}-2k^{3}(1+2w)+k^{2}(7+2w+6w^{2})\\
 & - & 2k(3+5w-w^{2}+2w^{3}))\sum_{i=0}^{w}\binom{i}{k}\Delta_{i}\\
 & + & \frac{1}{2}(1-k+w)^{2}(4+k^{2}+w^{2}-2k(1+w))\sum_{i=0}^{w}\binom{i}{k-1}\Delta_{i}\\
 & + & \frac{1}{4}(1-k+w)^{2}(2-k+w)^{2}\sum_{i=0}^{w}\binom{i}{k-2}\Delta_{i}\end{eqnarray*}
Note, that from this formula we can derive the expression for strength
of 2-perfect code in $J(2w,w)$ by substitution $k=\varphi+1.$ Hence,
we assume that $\sum\binom{i}{j}\Delta_{i}=0$ for all $j<k$, and
$\sum\binom{i}{k}\Delta_{i}\neq0$. Thus we obtain the following four
roots:

\begin{eqnarray*}
 & k=\frac{1}{2}\left(1+2w\mp\sqrt{-11+8w\mp4\sqrt{5-6w+2w^{2}}}\right)\end{eqnarray*}
or \[
\varphi=\frac{1}{2}\left(-1+2w\mp\sqrt{-11+8w\mp4\sqrt{5-6w+2w^{2}}}\right).\]
If we assume that $k=w-j+2$ we obtain the following recursion formula:

\[
\sum_{i=0}^{w}\binom{i}{w-j}\Delta_{i}=-\frac{F(w,j)\sum_{i=0}^{w}\binom{i}{w-j+2}\Delta_{i}+G(w,j)\sum_{i=0}^{w}\binom{i}{w-j+1}\Delta_{i}}{(j-1)^{2}j^{2}}\]
where $2\leq j<w-\varphi$, $F(w,j)=20+(j-3)j(10+(j-3)j)-14w-4(j-3)jw+2w^{2}$
and $G(w,j)=2(j-1)^{2}(4+(j-2)j-2w)$.

Since we consider the case $e=2$, we have two possibilities for $B_{i}$:
$B_{i,1}$ and $B_{i,2}$. In other words, we consider the number
of translate-words from configuration $(i,w-i)$ in the translate
with translate-leader $(w-1,1)$ and $(w-2,2)$, respectively.

Thus we have two possible $\Delta_{i}$: $\Delta_{i,1}$ and $\Delta_{i,2}$.

Now we compute binomial moments for the first several values of $j$.

From $\Delta_{w,l}=1,$ for $l=1,2$, we have $\sum_{i=0}^{w}\binom{i}{w}\Delta_{i,l}=\Delta_{w,l}=1.$

From $\Delta_{w-1,1}=-1$, $\Delta_{w-1,2}=0$, it follows: \[
\sum_{i=0}^{w}\binom{i}{w-1}\Delta_{i,l}=\binom{w-1}{w-1}\Delta_{w-1,l}+\binom{w}{w-1}\Delta_{w,l}=\left\{ \begin{array}{c}
w-1,\: l=1\\
w,\:\:\:\;\;\;\: l=2\end{array}\right.\]

\begin{itemize}
\item $j=2$. \begin{align*}
\sum_{i=0}^{w}\binom{i}{w-2}\Delta_{i,1} & =\frac{(w-1)(w-2)}{2}\\
\sum_{i=0}^{w}\binom{i}{w-2}\Delta_{i,2} & =\frac{(w+1)(w-2)}{2}\end{align*}

\item $j=3$.\begin{align*}
\sum_{i=0}^{w}\binom{i}{w-3}\Delta_{i,1} & =\frac{(w-1)(w-2)(w-3)}{6}\\
\sum_{i=0}^{w}\binom{i}{w-3}\Delta_{i,2} & =\frac{(w-2)(3w^{2}-5w-14)}{2*3^{2}}\end{align*}

\item $j=4$.\begin{align*}
\sum_{i=0}^{w}\binom{i}{w-4}\Delta_{i,1} & =\frac{(w-1)(w-2)(w-5)(5w-14)}{3^{2}4^{2}}\\
\sum_{i=0}^{w}\binom{i}{w-4}\Delta_{i,2} & =\frac{(w-2)(w-5)(5w^{2}-7w-26)}{3^{2}4^{2}}\end{align*}

\item $j=5$.\begin{align*}
\sum_{i=0}^{w}\binom{i}{w-5}\Delta_{i,1} & =\frac{(w-1)(w-2)(w-5)(334-171w+7w^{2})}{3^{2}4^{2}5^{2}}\\
\sum_{i=0}^{w}\binom{i}{w-5}\Delta_{i,2} & =\frac{(w-2)(w-5)(17w^{3}-147w^{2}+66w+680)}{3^{2}4^{2}5^{2}}.\end{align*}

\end{itemize}
Note that by \cite{Schwartz}, if $2$-perfect code exists in $J(2w,w)$,
then $w\equiv2,\:26\mbox{ \mbox{or }}50(\mbox{mod }60)$. But for
$w\equiv26(\mbox{mod }60)$ the last divisibility condition is not
satisfied, therefore remains only $w\equiv2\mbox{ or }50(\mbox{mod }60)$.

\begin{itemize}
\item $j=6$.\begin{align*}
\sum_{i=0}^{w}\binom{i}{w-6}\Delta_{i,1} & =\frac{2(w-1)(w-2)(w-5)(-5684+3544w-589w^{2}+29w^{3})}{3^{2}4^{2}5^{2}6^{2}}\\
\sum_{i=0}^{w}\binom{i}{w-6}\Delta_{i,2} & =\frac{2(w-2)(w-5)(-12228+228w+2663w^{2}-548w^{3}+29w^{4})}{3^{2}4^{2}5^{2}6^{2}}.\end{align*}

\item $j=7.$
\end{itemize}
\begin{eqnarray*}
\sum_{i=0}^{w}\binom{i}{w-7}\Delta_{i,1} & = & \frac{2(w-1)(w-2)(w-5)}{3^{2}4^{2}5^{2}6^{2}7^{2}}\\
 & * & (262324-185444w+39797w^{2}-3376w^{3}+99w^{4})\\
\sum_{i=0}^{w}\binom{i}{w-7}\Delta_{i,2} & = & \frac{2(w-2)(w-5)}{3^{2}4^{2}5^{2}6^{2}7^{2}}\\
 & * & (585224-59628w-123650w^{2}+34855w^{3}-3236w^{4}+99w^{5}).\end{eqnarray*}

The last divisibility conditions leave only the following values of
$w$ modulo $420$:

\begin{itemize}
\item $w\equiv2,\:302\mbox{ or }362(\mbox{mod }420);$
\item $w\equiv50,\:110\mbox{ or }170(\mbox{mod }420).$
\end{itemize}

\subsubsection{Necessary conditions for the existence of a 2-perfect code in $J(2w,w)$}

In this section we show the necessary conditions for the existence
of a 2-perfect code in $J(2w,w)$ using Pell equation and prove that
there are no 2-perfect codes in $J(2w,w)$ for $n<2.5*10^{15}$.

Assume $C$ is a 2-perfect code in $J(2w,\; w)$.

We saw that the strength of the code is:

\begin{eqnarray*}
\frac{1}{2}(-1+2w-\sqrt{8w-11\pm4\sqrt{5-6w+2w^{2}}}).\end{eqnarray*}
Hence, the first constraint is:

\[
\sqrt{5-6w+2w^{2}}\in\mathbb{Z}\]
therefore, $\exists y\in\mathbb{Z}$, s.t.\begin{eqnarray*}
5-6w+2w^{2} & = & y^{2}\\
10-12w+4w^{2} & = & 2y^{2}\\
(2w-3)^{2}-2y^{2} & = & -1\end{eqnarray*}
Let $x=2w-3$. This brings us to the Pell equation:\[
x^{2}-2y^{2}=-1\]
with the family of solutions in the form of:

\begin{eqnarray}
x & = & \frac{(1+\sqrt{2})^{k}+(1-\sqrt{2})^{k}}{2}\label{eq:5}\\
y & = & \frac{(1+\sqrt{2})^{k}-(1-\sqrt{2})^{k}}{2\sqrt{2}}\label{eq:6}\end{eqnarray}
where $k$ is odd \cite{Pell}.

Using the binomial formula, from (\ref{eq:5}) and denoting $k=2m+1$
we derive the following expression for $x$:

\begin{eqnarray*}
x & = & \frac{1}{2}[\sum_{i=0}^{2m+1}\binom{2m+1}{i}2^{\frac{i}{2}}+\sum_{i=0}^{2m+1}\binom{2m+1}{i}2^{\frac{i}{2}}(-1)^{i}]\\
 & = & \sum_{i\, is\: even}\binom{2m+1}{i}2^{\frac{i}{2}}=\sum_{j=0}^{m}\binom{2m+1}{2j}2^{j}\end{eqnarray*}
or

\begin{eqnarray*}
x & = & 1+\binom{2m+1}{2}2+\binom{2m+1}{4}2^{2}+...+\binom{2m+1}{2m}2^{m}.\end{eqnarray*}

We know from \cite{Schwartz} that if a $2$-perfect code exists in
$J(2w,w)$, then $w\equiv2,26,50(\mbox{mod }60)$, and thus $w\equiv2(\mbox{mod }12)$.

Since $w=\frac{x+3}{2}$, then $\exists z$, s.t. $12\; z=w-2=\frac{x+3}{2}-2=\frac{x-1}{2}$.
Consequently $24z=x-1$, $x\equiv1(\mbox{mod }24)$, and in particular,
$x\equiv1(\mbox{mod }4)$ and $x\equiv1(\mbox{mod }3)$.

\begin{itemize}
\item Since $x\equiv1(\mbox{mod }4)$ we have:
\end{itemize}
\[
1+\binom{2m+1}{2}2\equiv1(\mbox{mod }4)\]
or

\begin{eqnarray*}
2m(2m+1) & \equiv0(\mbox{mod }4)\end{eqnarray*}
therefore $m$ is even. Denote $m=2t$.

\begin{itemize}
\item Since $2\equiv-1(\mbox{mod }3)$, we have:
\end{itemize}
\[
2^{j}\equiv\left\{ \begin{array}{c}
2,\: j\mbox{\, is\, odd}\\
1,\: j\,\mbox{ is\, even}\end{array}\right.(\mbox{mod }3)\]
therefore, from $x\equiv1(\mbox{mod }3)$:\begin{eqnarray*}
\sum_{j\, is\, even}\binom{2m+1}{2j}+2\sum_{j\, is\, odd}\binom{2m+1}{2j} & \equiv & 0(\mbox{mod }3)\end{eqnarray*}
or\begin{eqnarray*}
\sum_{j\, is\, even}\binom{2m+1}{2j}-\sum_{j\, is\, odd}\binom{2m+1}{2j} & \equiv & 0(\mbox{mod }3)\end{eqnarray*}

For example, for $m=6,$ we obtain the contradiction:\begin{eqnarray*}
[\binom{13}{2}+\binom{13}{6}+\binom{13}{10}]-[\binom{13}{4}+\binom{13}{8}+\binom{13}{12}] & = & 65\neq0(\mbox{mod }3).\end{eqnarray*}
The second constraint is:\[
\sqrt{8w-11\pm4\sqrt{5-6w+2w^{2}}}\in\mathbb{Z}.\]
We examine two cases, positive root and negative root.

\begin{itemize}
\item $\sqrt{8w-11+4\sqrt{5-6w+2w^{2}}}\in\mathbb{Z}.$\begin{eqnarray*}
8w-11+4\sqrt{5-6w+2w^{2}} & = & 8w-11+4y\\
 & = & 8(\frac{x+3}{2})-11+4y\\
 & = & 4x+1+4y=4(x+y)+1\end{eqnarray*}
therefore, $\exists c\in\mathbb{Z}$, s.t.\[
4(x+y)+1=c^{2}\]

\item $\sqrt{8w-11-4\sqrt{5-6w+2w^{2}}}\in\mathbb{Z}.$
\end{itemize}
\begin{eqnarray*}
8w-11-4\sqrt{5-6w+2w^{2}} & = & 8w-11-4y\\
 & = & 8(\frac{x+3}{2})-11-4y\\
 & = & 4x+1-4y=4(x-y)+1\end{eqnarray*}

therefore, $\exists d\in\mathbb{Z}$, s.t.\[
4(x-y)+1=d^{2}\]
From (\ref{eq:5}) and (\ref{eq:6}) we obtain:

\begin{eqnarray}
x+y & = & \frac{\sqrt{2}(1+\sqrt{2})^{k}+\sqrt{2}(1-\sqrt{2})^{k}+(1+\sqrt{2})^{k}-(1-\sqrt{2})^{k}}{2\sqrt{2}}\nonumber \\
 & = & \frac{(\sqrt{2}+1)(1+\sqrt{2})^{k}+(\sqrt{2}-1)(1-\sqrt{2})^{k}}{2\sqrt{2}}\nonumber \\
 & = & \frac{(1+\sqrt{2})^{k+1}-(1-\sqrt{2})^{k+1}}{2\sqrt{2}}\label{eq:7}\end{eqnarray}

\begin{eqnarray}
x-y & = & \frac{\sqrt{2}(1+\sqrt{2})^{k}+\sqrt{2}(1-\sqrt{2})^{k}-(1+\sqrt{2})^{k}+(1-\sqrt{2})^{k}}{2\sqrt{2}}\nonumber \\
 & = & \frac{(\sqrt{2}-1)(\sqrt{2}+1)^{k}-(\sqrt{2}+1)(\sqrt{2}-1)^{k}}{2\sqrt{2}}\nonumber \\
 & = & \frac{(\sqrt{2}+1)^{k-1}-(\sqrt{2}-1)^{k-1}}{2\sqrt{2}}\label{eq:8}\end{eqnarray}
$k=2m+1$, $m=2t$ , thus we can substitute $k=4t+1$, and using the
binomial formula we have

\begin{eqnarray*}
x+y & = & \frac{1}{2\sqrt{2}}[\sum_{i=0}^{4t+2}\binom{4t+2}{i}2^{\frac{i}{2}}-\sum_{i=0}^{4t+2}\binom{4t+2}{i}2^{\frac{i}{2}}(-1)^{i}]\\
 & = & \frac{1}{\sqrt{2}}\sum_{i\, is\, odd}\binom{4t+2}{i}2^{\frac{i}{2}}=\sum_{i\, is\, odd}\binom{4t+2}{i}2^{\frac{i-1}{2}}\end{eqnarray*}
therefore,

\begin{eqnarray*}
4(x+y)+1 & = & 1+\sum_{i\, is\, odd}\binom{4t+2}{i}2^{\frac{i+3}{2}}\end{eqnarray*}
or denoting $i=2j+1$we can write\[
c^{2}=1+\sum_{j=0}^{2t}\binom{4t+2}{2j+1}2^{j+2}\]
We get the same result in the second case, too:

\begin{eqnarray*}
x-y & = & \frac{1}{2\sqrt{2}}[\sum_{i=0}^{4t}\binom{4t}{i}2^{\frac{i}{2}}-\sum_{i=0}^{4t}\binom{4t}{i}2^{\frac{i}{2}}(-1)^{i}]\\
 & = & \frac{1}{\sqrt{2}}\sum_{i\, is\, odd}\binom{4t}{i}2^{\frac{i}{2}}=\sum_{i\, is\, odd}\binom{4t}{i}2^{\frac{i-1}{2}}\end{eqnarray*}
therefore,

\begin{eqnarray*}
4(x-y)+1 & = & 1+\sum_{i\, is\, odd}\binom{4t}{i}2^{\frac{i+3}{2}}\end{eqnarray*}
or denoting $i=2j+1$ we can write:\[
d^{2}=1+\sum_{j=0}^{2t-1}\binom{4t}{2j+1}2^{j+2}.\]
We examine a few first values of $t$:

\begin{itemize}
\item for $t=0$ we have $c^{2}=9$, $d^{2}=1$;
\item for $t=1$ we have $c^{2}=281$, $d^{2}=49$, contradiction for $c$;
\item for $t=2$ we have $c^{2}=9513$, $d^{2}=1633$, contradiction.
\item for $t=3$ we have $c^{2}=323129$, $d^{2}=55441,$contradiction.
\end{itemize}
Now from (\ref{eq:7}), (\ref{eq:8}) and $k=4t+1$ we get\begin{eqnarray*}
c^{2}=1+4(x+y) & = & 1+4(\frac{(1+\sqrt{2})^{4t+2}-(1-\sqrt{2})^{4t+2}}{2\sqrt{2}})\\
 & = & 1+\sqrt{2}[((1+\sqrt{2})^{2})^{2t+1}-((1-\sqrt{2})^{2})^{2t+1}]\\
 & = & 1+\sqrt{2}[(3+2\sqrt{2})^{2t+1}-(3-2\sqrt{2})^{2t+1}]\end{eqnarray*}

\begin{eqnarray*}
d^{2}=1+4(x-y) & = & 1+4(\frac{(\sqrt{2}+1)^{4t}-(\sqrt{2}-1)^{4t}}{2\sqrt{2}})\\
 & = & 1+\sqrt{2}[((\sqrt{2}+1)^{2})^{2t}-((\sqrt{2}-1)^{2})^{2t}]\\
 & = & 1+\sqrt{2}[(3+2\sqrt{2})^{2t}-(3-2\sqrt{2})^{2t}]\\
 & = & 1+\sqrt{2}[(17+12\sqrt{2})^{t}-(17-12\sqrt{2})^{t}]\end{eqnarray*}

Using these expressions above we build the following Table \ref{tab:pell}
for several $t$: (recall that $k=4t+1,$ where $k$ is the exponent
in the expression for $x$ and $y$).

\begin{table}[H]
\caption{\label{tab:pell}nonexistence of $2$-perfect codes in $J(2w,w)$
for $n<2.5*10^{15}$}

\begin{tabular}{|c|c|c|c|c|}
\hline
$t$ & $1+4(x-y)$ & $1+4(x+y)$ & $x$ & $w=\frac{x+3}{2}$\tabularnewline
\hline
\hline
0 & 1 & 9 & 1 & 2\tabularnewline
\hline
1 & 49 & 281 & 41 & 22$(\neq2(12)$\tabularnewline
\hline
2 & 1633 & 9513 & 1393 & .\tabularnewline
\hline
3 & 55441 & 323129 & 47321 & .\tabularnewline
\hline
4 & 1883329 & 10976841 & 1607521 & .\tabularnewline
\hline
5 & 63977713 & 372889433 & . & .\tabularnewline
\hline
6 & 2173358881 & 12667263849 & . & .\tabularnewline
\hline
7 & 73830224209 & 430314081401 & . & .\tabularnewline
\hline
8 & 2508054264193 & 14618011503753 & . & 1070379110498\tabularnewline
\hline
9 & 85200014758321 & 496582077046169 & . & 36361380737782\tabularnewline
\hline
10 & 2894292447518689 & 16869172608065961 & . & 1235216565974042\tabularnewline
\hline
\end{tabular}
\end{table}

Therefore, at least for $n<2.5*10^{15}$, the necessary condition
is not satisfied.

\noindent \textbf{Conclusion:} from the fact that two roots in the
expression for a code strength must be integers and the fact that
$x\equiv1(4)$, we prove that there is no 2-perfect code in $J(n,w)$
where $n=2w$ , for $n<2.5*10^{15}$.

In summary, we proved the following theorem:

\noindent \textbf{Theorem 50.} If $2$-perfect code $C$ exists in
$J(2w,w)$ then

\begin{enumerate}
\item $w=\frac{(1+\sqrt{2})^{4t+1}+(1-\sqrt{2})^{4t+1}+6}{4}$, for some
integer $t$.
\item $\sum_{j\;\mbox{is even}}\binom{4t+1}{2j}-\sum_{j\,\mbox{is\, odd}}\binom{4t+1}{2j}\equiv0(\mbox{mod }3)$.${\scriptstyle }$
\item $1+\sum_{j=0}^{2t}\binom{4t+2}{2j+1}2^{j+2}=1+\sqrt{2}[(3+2\sqrt{2})^{2t+1}-(3-2\sqrt{2})^{2t+1}]$
must be square of integer, if the strength of $C$ is $\sqrt{8w-11+4\sqrt{5-6w+2w^{2}}}$.
\item $1+\sum_{j=0}^{2t-1}\binom{4t}{2j+1}2^{j+2}=1+\sqrt{2}[(17+12\sqrt{2})^{t}-(17-12\sqrt{2})^{t}]$
must be square of integer, if the strength of $C$ is $\sqrt{8w-11-4\sqrt{5-6w+2w^{2}}}$.
\end{enumerate}
\hfill{}$\square$

\subsection{Binomial moments for $e-$perfect code in $J(2w,w)$.}

In this section we obtain the expression for $k$-th binomial moments
with respect to the difference configuration distributions for $k\geq\varphi+1$,
where $\varphi$ is the strength of an $e-$perfect code in $J(2w,w).$

Let $C$ be an $e$-perfect code in $J(2w,w)$. Let $\left\{ \alpha_{1},\alpha_{2}\right\} $
be a partition of $N$ such that $|\alpha_{1}|=w$, $|\alpha_{2}|=w$,
and a vector of $(w,0)$ configuration be a codeword. Let $A_{i}$
be the number of codewords from configuration $(w-i,i)$.

Let $B_{i}$ be the number of codewords from configuration $(w-i,i)$
in the translate with translate-leader $(w-1,1)$.

Let $\left\{ H_{1},H_{2},H_{3},H_{4}\right\} $ be a partition of
coordinate set $N$ with $|H_{1}|=|H_{4}|=w-1$, $|H_{2}|=|H_{3}|=1$
such that $H_{1}\cup H_{2}=\alpha_{1}$ and $H_{3}\cup H_{4}=\alpha_{2},$
and let

\begin{itemize}
\item $A_{01}^{i}=\frac{i^{2}}{w^{2}}A_{i}$ the number of codewords from
configuration $(w-i,0,1,i-1)$,
\item $A_{10}^{i}=\frac{(w-i)^{2}}{w^{2}}A_{i}$ the number of codewords
from configuration $(w-i-1,1,0,i)$,
\item $A_{00}^{i}=\frac{(w-i)i}{w^{2}}A_{i}$ the number of codewords from
configuration $(w-i,0,0,i)$,
\item $A_{11}^{i}=\frac{(w-i)i}{w^{2}}A_{i}$ the number of codewords from
configuration $(w-i-1,1,1,i-1)$.
\end{itemize}
Note that \[
A_{i}=A_{01}^{i}+A_{10}^{i}+A_{00}^{i}+A_{11}^{i}\]

\begin{eqnarray*}
B_{i} & = & A_{10}^{i-1}+A_{01}^{i+1}+A_{00}^{i}+A_{11}^{i}\end{eqnarray*}

Let $\varphi$ be the strength of the code. By Lemma 44 for $k\leq\varphi$
we have: \[
\sum_{i=0}^{w}\binom{i}{k}A_{i}=\sum_{i=0}^{w}\binom{i}{k}B_{i}=\binom{w}{k}\frac{\binom{n-k}{w-k}}{\Phi_{e}(n,w)}=\frac{|C|\binom{w}{k}}{\binom{n}{k}}\binom{w}{k}.\]

\noindent \textbf{Theorem} \textbf{51}. If $C$ is an $e$-perfect
code in $J(2w,w)$ and $\varphi$ is its strength, then for each $k$,
$\varphi<k\leq w$, we have

\begin{eqnarray*}
w^{2}\sum_{i=0}^{w}\binom{i}{k}\Delta_{i} & = & (2wk-k^{2}+k)\sum_{i=0}^{w}\binom{i}{k}A_{i}-(w-k+1)^{2}\sum_{i=0}^{w}\binom{i}{k-1}A_{i}.\end{eqnarray*}

\noindent \emph{Proof}. For $k\geq\varphi+1$ we have:

\begin{eqnarray*}
|C|\binom{w}{k} & = & 2\sum_{i=0}^{w}\binom{i}{k}A_{i}+X,\\
|C|\binom{w}{k} & = & 2\sum_{i=0}^{w}\binom{i}{k}B_{i}+Y,\end{eqnarray*}
where left part of the equations is the number of ways to choose $k$
columns, and the first summand of the right part is the number of
ways to choose $k$ columns in only one part of $w$ coordinates,
and the second summand of the right part is the number of ways to
choose $k$ columns which appear in more than one part of $w$ coordinates.

\begin{eqnarray*}
X & = & \sum_{i=0}^{w}(X_{01}^{i}+X_{10}^{i}+X_{00}^{i}+X_{11}^{i})\\
Y & = & \sum_{i=0}^{w}(Y_{01}^{i}+Y_{10}^{i}+Y_{00}^{i}+Y_{11}^{i})\end{eqnarray*}
where \begin{eqnarray*}
X_{lj}^{i} & = & [\binom{w}{k}-\binom{i}{k}-\binom{w-i}{k}]A_{lj}^{i}\end{eqnarray*}

\begin{eqnarray*}
Y_{01}^{i} & = & [\binom{w}{k}-\binom{i-1}{k}-\binom{w-i+1}{k}]A_{01}^{i}\\
Y_{10}^{i} & = & [\binom{w}{k}-\binom{i+1}{k}-\binom{w-i-1}{k}]A_{10}^{i}\\
Y_{00}^{i} & = & X_{00}^{i}\\
Y_{11}^{i} & = & X_{11}^{i}\end{eqnarray*}
since $\Delta_{i}=A_{i}-B_{i}$, we get:\begin{align*}
0 & =2\sum_{i=0}^{w}\binom{i}{k}\Delta_{i}+\sum_{i=0}^{w}[\binom{w}{k}-\binom{i}{k}-\binom{w-i}{k}]A_{01}^{i}+\sum_{i=0}^{w}[\binom{w}{k}-\binom{i}{k}-\binom{w-i}{k}]A_{10}^{i}\\
 & -\sum_{i=0}^{w}[\binom{w}{k}-\binom{i-1}{k}-\binom{w-i+1}{k}]A_{01}^{i}-\sum_{i=0}^{w}[\binom{w}{k}-\binom{i+1}{k}-\binom{w-i-1}{k}]A_{10}^{i}\end{align*}
We substitute the expressions for $A_{01}^{i}$ and $A_{10}^{i}$
:\begin{align*}
0 & =2\sum_{i=0}^{w}\binom{i}{k}\Delta_{i}+\sum_{i=0}^{w}[\binom{i-1}{k}+\binom{w-i+1}{k}-\binom{i}{k}-\binom{w-i}{k}]\frac{i^{2}}{w^{2}}A_{i}\\
 & +\sum_{i=0}^{w}[\binom{i+1}{k}+\binom{w-i-1}{k}-\binom{i}{k}-\binom{w-i}{k}]\frac{(w-i)^{2}}{w^{2}}A_{i},\end{align*}
or \begin{eqnarray*}
2w^{2}\sum_{i=0}^{w}\binom{i}{k}\Delta_{i} & = & \sum_{i=0}^{w}[\binom{w-i-1}{k-1}(w-i)^{2}-\binom{w-i}{k-1}i^{2}\\
 & + & \binom{i-1}{k-1}i^{2}-\binom{i}{k-1}(w-i)^{2}]A_{i}\end{eqnarray*}
Using the fact that the code is self-complement, we prove the following
proposition (see Appendix C).

\noindent \textbf{Proposition} \textbf{52}.\[
\sum_{i=0}^{w}[\binom{w-i-1}{k-1}(w-i)^{2}-\binom{w-i}{k-1}i^{2}+\binom{i-1}{k-1}i^{2}-\binom{i}{k-1}(w-i)^{2}]A_{i}=\]
\[
=2(2wk-k^{2}+k)\sum_{i=0}^{w}\binom{i}{k}A_{i}-2(w-(k-1))^{2}\sum_{i=0}^{w}\binom{i}{k-1}A_{i}.\]
Therefore, we have that \[
w^{2}\sum_{i=0}^{w}\binom{i}{k}\Delta_{i}=(2wk-k^{2}+k)\sum_{i=0}^{w}\binom{i}{k}A_{i}-(w-(k-1))^{2}\sum_{i=0}^{w}\binom{i}{k-1}A_{i}.\]

\hfill{}$\square$

\chapter{Perfect doubly constant weight codes}

Constant weight codes are building blocks for general codes in Hamming
metric. Similarly, doubly constant weight are building blocks for
codes in Johnson metric. Doubly constant weight codes play an important
role in obtaining bounds on the sizes of constant weight codes. A
natural question is whether there exist perfect doubly constant weight
codes.

In this chapter we discuss three types of trivial perfect doubly constant
weight codes, show some properties of perfect doubly constant weight
codes, construct the family of parameters for  codes whose sphere
divides the size of whole space (while in Johnson graph we do not
know codes with such parameters), and present the necessary condition
for existence of an e-perfect code, which is equivalent to Roos' bound
in Johnson graph.

\section{Definitions and properties of perfect doubly constant weight codes}

Given five integers, $n_{1},n_{2},w_{1},w_{2}$ and $d$, such that
$0\leq w_{1}\leq n_{1}$ and $0\leq w_{2}\leq n_{2}$, define \emph{doubly
constant weight code} $(w_{1},n_{1},w_{2},n_{2},d)$ be a constant
weight code of length $n_{1}+n_{2}$ and weight $w_{1}+w_{2}$, with
$w_{1}$ ones in the first $n_{1}$ positions and $w_{2}$ ones in
the last $n_{2}$ positions, and minimum distance $d$. Note, that
because this definition is based on the definition of constant weight
codes, the distance $d$ denotes J-distance, as before.

Let $T(w_{1},n_{1},w_{2},n_{2},\delta)$ denote the maximum number
of codewords in a $(w_{1},n_{1},w_{2},$ $n_{2},d)$ code, where $\delta=2d$
is a H-distance. Upper bounds on $T(w_{1},n_{1},w_{2},n_{2},\delta)$
were found and used in \cite{Agrell} to find upper bounds on $A(n,\delta',w)$.

We denote as $V_{w_{1},w_{2}}^{n_{1},n_{2}}$ the space of all binary
vectors of length $n_{1}+n_{2}$ and weight $w_{1}+w_{2},$ with $w_{1}$
ones in the first $n_{1}$ positions and $w_{2}$ ones in the last
$n_{2}$ positions.

A doubly constant weight code $C$ is called an $e$-\emph{perfect
code,} if the $e$-spheres of all the codewords of $C$ form a partition
of $V_{w_{1},w_{2}}^{n_{1},n_{2}}$.

The number of codewords of an $e$-perfect code $C=(w_{1},n_{1},w_{2},n_{2},d)$
is \[
|C|=\frac{\binom{n_{1}}{w_{1}}\binom{n_{2}}{w_{2}}}{\Phi_{e}(n_{1},w_{1},n_{2},w_{2})}\]
where \[
\Phi_{e}(n_{1},w_{1},n_{2},w_{2})=\sum_{i=0}^{e}\sum_{j=0}^{e-i}\binom{w_{1}}{i}\binom{n_{1}-w_{1}}{i}\binom{w_{2}}{j}\binom{n_{2}-w_{2}}{j},\]
and hence we have that\begin{equation}
\Phi_{e}(n_{1},w_{1},n_{2},w_{2})\mid\binom{n_{1}}{w_{1}}\binom{n_{2}}{w_{2}}.\label{eq:9}\end{equation}

There are some trivial perfect doubly constant weight codes:

\begin{enumerate}
\item $V_{w_{1},w_{2}}^{n_{1},n_{2}}$ is $0$-perfect.
\item Any $\left\{ v\right\} $, $v\in V_{w_{1},w_{2}}^{n_{1},n_{2}}$,
is $(w_{1}$$+w_{2})$-perfect.
\item If $n_{1}=2w_{1}$, $n_{2}=2w_{2}$ and $w_{1}+w_{2}$ is odd, then
any pair of vectors with disjoint $w_{1}+w_{2}$ sets of ones (with
$w_{1}$ ones in the first $n_{1}$ positions and $w_{2}$ ones in
the last $n_{2}$ positions) is $e$-perfect with $e=\frac{w_{1}+w_{2}-1}{2}$.
\end{enumerate}
\textbf{Lemma 53.} If $C$ is an $e$-perfect doubly constant weight
code then its minimum J-distance is $2e+1$.

\noindent \emph{Proof}. Since $C$ is an $e$-perfect code, it follows
that the $e$-spheres of two codewords with J-distance less than $2e+1$
have nonempty intersection. Hence, the minimum J-distance of the code
is $2e+1$.

\hfill{}$\square$

\noindent \textbf{Lemma 54.} If $C$ is an $e$-perfect doubly constant
weight code then $T(w_{1},n_{1},w_{2},n_{2},4e+2)=|C|$.

\noindent \emph{Proof}\textbf{.} Assume $C$ is an $e$-perfect doubly
constant weight code, then by Lemma 53, it is $(w_{1},n_{1},w_{2},n_{2},2e+1)$
code and hence the $e$-spheres around its codewords are disjoint.
Since all $e$-spheres have the same size and they form partition
of $V_{w_{1},w_{2}}^{n_{1},n_{2}}$, then $T(w_{1},n_{1},w_{2},n_{2},4e+2)=|C|$.

\hfill{}$\square$

\noindent \textbf{Lemma 55.} If $C=(w_{1},n_{1},w_{2},n_{2},2e+1)$
is an $e$-perfect doubly constant weight code then the complement
of $C$ in the first $n_{1}$ positions is an e-perfect code $(n_{1}-w_{1},n_{1},w_{2},n_{2},2e+1)$.

\noindent \emph{Proof}\textbf{.} The Lemma follows from the fact that
there exists an isomorphism between the space of all binary vectors
of length $n_{1}+n_{2}$ and weight $w_{1}+w_{2},$ with $w_{1}$
ones in the first $n_{1}$ positions and $w_{2}$ ones in the last
$n_{2}$ positions and its complement in the first $n_{1}$ positions.

\hfill{}$\square$

\noindent \textbf{Corollary 56.} If $C=(w_{1},n_{1},w_{2},n_{2},2e+1)$
is an $e$-perfect doubly constant weight code then the complement
of $C$ in the last $n_{2}$ positions is an e-perfect code $(w_{1},n_{1},n_{2}-w_{2},n_{2},2e+1)$.

\noindent \textbf{Corollary 57.} If $C=(w_{1},n_{1},w_{2},n_{2},2e+1)$
is an $e$-perfect doubly constant weight code then the complement
of $C$ is an e-perfect code $(n_{1}-w_{1},n_{1},n_{2}-w_{2},n_{2},2e+1)$.

From Lemma 53 , Lemma 55 ,Corollary 56 and Corollary 57  follows:

\noindent \textbf{Corollary 58.} If $C=(w_{1},n_{1},w_{2},n_{2},2e+1)$
is a non trivial $e$-perfect doubly constant weight code then $w_{1}+w_{2}\geq2e+1$
, $n_{1}+n_{2}-w_{1}-w_{2}\geq2e+1$, $n_{1}-w_{1}+w_{2}\geq2e+1$
and $w_{1}+n_{2}-w_{2}\geq2e+1$.

\section{Family of parameters for codes whose size of sphere, $\Phi_{1}(n_{1},w_{1},n_{2},w_{2})$,
divides the size of whole space}

In this section we show the family of parameters for codes that satisfy
the necessary condition (\ref{eq:9}) for existence a 1-perfect doubly
constant weight code.

\noindent \textbf{Proposition 59.} Let $k$ be a natural number and
$C$ be a doubly constant weight code $(w_{1},n_{1},w_{2},n_{2},3)$,
when $w_{1}=w_{2}=2k$, $n_{1}=4k+1$, and $n_{2}=4k+2$. Then $\Phi_{1}(n_{1},w_{1},n_{2},w_{2})\mid\binom{n_{1}}{w_{1}}\binom{n_{2}}{w_{2}}$.

\noindent \emph{Proof}\textbf{.} \begin{align*}
\Phi_{1}(n_{1},w_{1},n_{2},w_{2}) & =1+w_{1}(n_{1}-w_{1})+w_{2}(n_{2}-w_{2})\\
 & =1+2k(2k+1)+2k(2k+2)=(2k+1)(4k+1),\end{align*}
therefore we have to prove that \[
\frac{\binom{4k+1}{2k}\binom{4k+2}{2k}}{(2k+1)(4k+1)}\in\mathbb{Z}.\]
But \[
\frac{\binom{4k+1}{2k}}{4k+1}=\binom{4k}{2k}\frac{1}{2k+1}\in\mathbb{Z}\]
is a Catalan number \cite{catalan} and

\[
\frac{\binom{4k+2}{2k}}{2k+1}=\binom{4k+2}{2k+1}\frac{1}{2k+2}\in\mathbb{Z}\]
is also Catalan number, hence \[
\Phi_{1}(4k+1,2k,4k+2,2k)\mid\binom{4k+1}{2k}\binom{4k+2}{2k}.\]

\hfill{}$\square$

Codes with parameters as above are candidates for being perfect codes.
But from \cite{Tables} we can see that for small $k$ ($k=1,2,$or
$3$) there are no $1-$perfect doubly constant weight codes with
such parameters. Still, we can not say anything about the codes with
higher values of $k$.

\section{Necessary condition for existence of an $e$-perfect doubly constant
weight code}

In this section we prove the theorem that gives the bound for parameters
of $e$-perfect code. This bound is similar to the Roos's bound in
Johnson graph. Hence, the techniques that we use here are a generalization
of the ideas of the proof of Roos' bound by Etzion \cite{on_perfect}.

We recall a few definitions which we will use in the proof of the
existence theorem.

For a given partition of set of all $n_{1}+n_{2}$ coordinates into
four subsets $\alpha,\beta,\gamma$ and $\delta$, let \emph{configuration}
$(a,b,c,d)$ be a set of all vectors with weight $a$ in the positions
of $\alpha$, weight $b$ in the positions of $\beta$, weight $c$
in the positions of $\gamma$ and weight $d$ in the positions of
$\delta$.

For an $e$-perfect doubly constant weight code $C$ we say that $w\in C$
\emph{J-cover} $v\in V_{w_{1},w_{2}}^{n_{1},n_{2}}$ if the J-distance
between $u$ and $v$ less or equal to $e$.

\noindent \textbf{Theorem 60 .} If an $e$-perfect doubly constant
weight code $(w_{1},n_{1},w_{2},n_{2},2e+1)$ exists then\[
n_{1}\leq\frac{(2e+1)(w_{1}-1)+w_{2}}{e}\]
and\[
n_{2}\leq\frac{(2e+1)(w_{2}-1)+w_{1}}{e}.\]

\noindent \emph{Proof}. Assume $C$ is an $e$-perfect code $(w_{1},n_{1},w_{2},n_{2},2e+1)$.

\emph{Case 1}: $w_{1}>e$

We partition the set of coordinates into four subsets $\alpha,\beta,\gamma$
and $\delta$ such that $\mid\alpha\mid=w_{1}-1$, $\mid\beta\mid=w_{2}$,
$\mid\gamma\mid=n_{1}-w_{1}+1$, $\mid\delta\mid=n_{2}-w_{2}$, and
there is a codeword of configuration $(w_{1}-(e+1),w_{2},e+1,0)$.
The J-distance between a vector from configuration $(w_{1}-(e+1),w_{2},e+1,0)$
and a vector from configuration $(w_{1}-a,w_{2}-b,a,b)$, $0<a+b\leq e$,
is strictly less than $2e+1$, so $C$ does not have any codeword
from configuration $(w_{1}-a,w_{2}-b,a,b)$, $0<a+b\leq e$. Therefore,
all the vectors from configuration $(w_{1}-1,w_{2},1,0)$ are J-covered
by codewords from configuration $(w_{1}-(e+1),w_{2},e+1,0)$, or $(w_{1}-e,w_{2}-1,e,1)$,
or $(w_{1}-(e-1),w_{2}-2,e-1,2)$,$...,$ or $(w_{1}-1,w_{2}-e,1,e)$.

Let $X_{i}$, $0\leq i\leq e$, be a collection of codewords from
configuration $(w_{1}-(e+1-i),w_{2}-i,e+1-i,i)$, such that $\bigcup_{i=0}^{e}X_{i}$
$J-$cover all the vectors from configuration $(w_{1}-1,w_{2},1,0)$.
There are $n_{1}-w_{1}+1$ vectors from configuration $(w_{1}-1,w_{2},1,0)$
and each codeword in $X_{i}$ J-covers $e+1-i$ such vectors. Therefore,\begin{equation}
\sum_{i=0}^{e}(e+1-i)|X_{i}|=n_{1}-w_{1}+1.\label{eq:10}\end{equation}
Since the minimum J-distance is $2e+1$, two codewords in $\bigcup_{i=0}^{e-1}X_{i}$
cannot intersect in the zeroes of part $\alpha$, and two codewords
in $\bigcup_{i=1}^{e}X_{i}$ cannot intersect in the zeroes of part
$\beta$. Hence,\begin{eqnarray}
\sum_{i=0}^{e-1}(e-i)|X_{i}| & \leq & w_{1}-1\label{eq:11}\end{eqnarray}

\begin{equation}
\sum_{i=1}^{e}i|X_{i}|\leq w_{2}.\label{eq:12}\end{equation}
Since\[
\frac{e+1}{e}\sum_{i=0}^{e-1}(e-i)|X_{i}|+\frac{1}{e}\sum_{i=1}^{e}i|X_{i}|=\sum_{i=0}^{e}(e+1-i)|X_{i}|,\]
from (\ref{eq:10}), (\ref{eq:11}) and (\ref{eq:12} ) above follows:

\[
n_{1}-w_{1}+1=\sum_{i=0}^{e}(e+1-i)|X_{i}|=\frac{e+1}{e}\sum_{i=0}^{e-1}(e-i)|X_{i}|+\frac{1}{e}\sum_{i=1}^{e}i|X_{i}|\leq\frac{e+1}{e}(w_{1}-1)+\frac{1}{e}w_{2}\]
Therefore,

\[
n_{1}\leq(w_{1}-1)(\frac{e+1}{e}+1)+\frac{w_{2}}{e}=\frac{(2e+1)(w_{1}-1)+w_{2}}{e}.\]

\emph{Case 2:} $1<w_{1}\leq e.$

Let $w_{1}=e-k$ for some $k$, $0\leq k<e-1$.

We use the same partition as in the Case 1: we partition the set of
coordinates into four subsets $\alpha,\beta,\gamma$ and $\delta$
such that $\mid\alpha\mid=w_{1}-1=e-k-1$, $\mid\beta\mid=w_{2}$,
$\mid\gamma\mid=n_{1}-w_{1}+1=n_{1}-e+k+1$, $\mid\delta\mid=n_{2}-w_{2}$,
and there is a codeword of configuration $(0,w_{2}-k-1,e-k,k+1)$.
All the vectors from configuration $(e-k-1,w_{2},1,0)$ are J-covered
only by codewords from configuration $(0,w_{2}-k-1,e-k,k+1)$ or from
configuration $(e-k-1,w_{2}-e,1,e)$, because of the restriction on
minimal distance $2e+1$ .

Let $X$ be a set of codewords from configuration $(0,w_{2}-k-1,e-k,k+1)$
and $Y$ a set of codewords from configuration $(e-k-1,w_{2}-e,1,e)$,
such that codewords in $X\bigcup Y$ cover all the vectors from configuration
$(e-k-1,w_{2},1,0)$. Therefore,\begin{equation}
(e-k)|X|+|Y|=n_{1}-e+k+1.\label{eq:13}\end{equation}
Note, that the J-distance between two codewords in $X$ less or equal
then $e+k+2.$ As $e-1>k$, it follows that \begin{equation}
\mid X\mid\leq1,\label{eq:14}\end{equation}
Since the minimum J-distance is $2e+1$, two codewords in $X\bigcup Y$
cannot intersect in the zeroes of part $\beta$. Hence,\begin{equation}
|Y|\: e\leq w_{2}-k-1,\label{eq:15}\end{equation}
From (\ref{eq:13}), (\ref{eq:14}) and (\ref{eq:15}) follows\[
n_{1}-e+k+1\leq e-k+\frac{w_{2}-k-1}{e}.\]
Thus,\begin{eqnarray*}
n_{1} & \leq & e-k-1+e-k+\frac{w_{2}-k-1}{e}=\frac{(2e-2k-1)e+w_{2}-k-1}{e}\\
 & = & \frac{2e^{2}-2ke-e+w_{2}-k-1}{e}=\frac{(2e+1)(e-k-1)+w_{2}}{e}.\end{eqnarray*}

\emph{Case 3:} $w_{1}=1.$

Now our partition is as follows: $\mid\alpha\mid=0,$$\mid\beta\mid=w_{2},$
$\mid\gamma\mid=n_{1}$,$\mid\delta\mid=n_{2}-w_{2}$ and there is
a codeword of configuration $(0,w_{2}-e,1,e)$. Let $X$ be a set
of codewords from configuration $(0,w_{2}-e,1,e)$. Hence, all the
vectors from configuration $(0,w_{2},1,0)$ are J-covered by the codewords
from $X.$ In addition, two codewords in $X$ cannot intersect in
the zeroes part $\beta$. Therefore,\[
n_{1}=|X|\leq\frac{w_{2}}{e}=\frac{(2e+1)(w_{1}-1)+w_{2}}{e}.\]
As we can swap the roles of $n_{1}$ and $n_{2}$, and $w_{1}$ and
$w_{2}$ we obtain the bound on $n_{2}$:

\[
n_{2}\leq\frac{(2e+1)(w_{2}-1)+w_{1}}{e}.\]

\hfill{}$\square$

\chapter{Steiner Systems and doubly Steiner Systems}

There is tight connection between constant weight codes and Steiner
systems, and doubly constant weight codes and doubly Steiner systems.
As an example of such connections, observe Steiner systems which are
optimal constant weight codes and doubly Steiner Systems which are
optimal doubly weight codes \cite{doubly}.

This chapter is organized as follows. In Section 4.1 we give definitions
and theorems that will be used in the following sections. In Section
4.2 we prove the bound on the length of Steiner system using anticodes.
In Section 4.3 we consider the doubly Steiner system and get analogous
results in this structure.

\section{Definitions and known results}

Let us recall the definition of Steiner systems.

A Steiner System $S(t,w,n)$ is a collection of $w$-subsets (called
blocks) taken from an $n$-set such that each $t$-subset of the $n$-set
is contained in exactly one block.

If we represent blocks as 0-1
-vectors we observe that a Steiner system $S(t,w,n)$ is equivalent
to a constant weight code with parameters $(n,2(w-t+1),w)$, since
any two vectors have at most $t-1$ ones in common.

Steiner systems play an important role in ruling out the existence
of $e$-perfect codes in $J(n,w)$. Moreover, the Steiner systems
$S(1,w,2w)$, where $w$ is odd, and $S(w,w,n)$, are among the trivial
perfect codes in the Johnson graph. Etzion proved that there are no
more Steiner systems which are also perfect codes in the Johnson graph
\cite{nonexist}.

We remind a few definitions which we will use in the following.

A connected graph $\Gamma$ with diameter $d$ is called \emph{distance-regular}
if for any vertices $x$ and $y$ of $\Gamma$ and any integers $0\leq i$,
$j\leq d$ , the number of vertices $z$ at distance $i$ from $x$
and at distance $j$ from $y$ depends only on $i$, $j$ and $k:=\mbox{dist}(x,y)$
and not on the choice of $x$ and $y$ themselves.

The following theorem is due to Delsarte\cite{Delsarte}:

\noindent \textbf{Theorem 61} : Let $X$ and $Y$ be subsets of the
vertex set $V$ of a distance regular graph $\Gamma$, such that nonzero
distances occurring between vertex in $X$ do not occur between vertices
of $Y$. Then $\mid X\mid\cdot\mid Y\mid\leq\mid V\mid$.

A subset $X$ of $V$ is called an \emph{anticode} with diameter \textbf{$D$},
if $D$ is the maximum distance occurring between vertices of $X$.

Anticodes with diameter $D$ having maximal size are called \emph{optimal
anticodes}.

Ahlswede, Aydinian and Khachatrian \cite{Ahlswede} gave a new definition
of diameter-perfect codes ($D$-perfect codes). They examined a variant
of Theorem 61.

Let $\Gamma$ be a distance-regular graph with a vertex set $V$.
If $A$ is an anticode in $\Gamma$, denote by $D(A)$ the diameter
of $A$. Let $A^{*}(D)=$$\max\left\{ \mid A\mid:D(A)\leq D\right\} $.

\noindent \textbf{Theorem 62 \cite{Ahlswede}} . If $C$ is a code
in $\Gamma$ with minimum distance $D+1$, then $\mid C\mid\leq\mid V\mid\cdot(A^{*}(D{}^{-1}.$

A code $C$ with minimum distance $D+1$ is called $D$-perfect if
Theorem 62 holds with equality. This is a generalization of the usual
definition of $e-$perfect codes as $e$-spheres are anticodes with
diameter $2e$.

\noindent \textbf{Lemma} \textbf{63} \cite{Ahlswede}. Any Steiner
system $S(t,w,n)$ forms a diameter perfect code.

We show the proof from \cite{Ahlswede} for completeness, since we
use it in the next section.

\noindent \emph{Proof.} Let $C$ be an $(n,2(w-t+1),w)$- code corresponding
to a $S(t,w,n)$. Then \[
|C|=\frac{\binom{n}{t}}{\binom{w}{t}}=\frac{\binom{n}{w}}{\binom{n-t}{w-t}}.\]
On the other hand $|C|\leq\frac{\binom{n}{w}}{A^{*}(n,2(w-t),w)}$,
where $A^{*}(n,2(w-t),w)$ is an optimal anticode in $J(n,w)$ of
diameter $2(w-t)$ (H-distance). Therefore $A^{*}(n,2(w-t),w)\leq\binom{n-t}{w-t}$.
Since there exists an anticode of size $\binom{n-t}{w-t}$ the statement
follows.

\hfill{}$\square$

\section{Necessary condition for existence of Steiner system }

In this section we provide an anticode-based proof of the bound on
Steiner system, which is different from the existing proof of Tits
\cite{Tits}. We note that similar two techniques were used to prove
Roos's bound, one by Roos \cite{Roos} based on anticodes and the
Theorem 61 of Delsarte, and another one by Etzion \cite{on_perfect}
based on specific partition of set of coordinates and J-covering some
vectors by codeword of specific configuration.

We first mention the proof by Tits for completeness.

\noindent \textbf{Theorem 64.} If Steiner System $S(t,w,n)$ exists
with $w<n$ then \[
n\geq(t+1)(w-t+1).\]

\noindent \emph{Proof 1} (Tits 1964 \cite{Tits}):

Let $T$ be a $t+1$-subset of the $n$-set, such that $T\nsubseteq B$,
for all blocks $B$. Such a $t+1$-set $T$ exists. There exactly
$t+1$ blocks $B_{0},...,B_{t}$ with $\mid B_{i}\cap T\mid=t$ ($i=0,...,t$).
The point sets $B_{i}\setminus T$ are mutually disjoint. Hence\[
n\geq\mid T\mid+\sum_{i=0}^{t}\mid B_{i}\setminus T\mid=(w-t+1)(t+1)\]

\hfill{}$\square$

\noindent \emph{Proof 2} (based on anticodes):

Assume $S(t,w,n)$ exists. Then by Lemma 63 for any anticode $A(n,w-t,w)$
in $J(n,w)$ with diameter $w-t$ (J- distance) we have\begin{equation}
A(n,w-t,w)\leq\binom{n-t}{w-t},\label{eq:16}\end{equation}
since we know that there is an optimal anticode with diameter $w-t$
and size $\binom{n-t}{w-t}$.

We will construct an anticode with diameter $w-t$ for Steiner system
$S(t,w,n)$.

Let $S$ be a set of coordinates of size $t+2.$ Denote $A_{t}$ to
be a collection of sets of coordinates of size $w$ which intersects
the given set $S$ in at least $t+1$ coordinates. We get the anticode
with diameter $w-t$ and size $\binom{n-t-2}{w-t-2}+(t+2)\binom{n-t-2}{w-t-1}$.
From (\ref{eq:16}) we have\[
\binom{n-t-2}{w-t-2}+(t+2)\binom{n-t-2}{w-t-1}\leq\binom{n-t}{w-t},\]
or \[
n\geq(t+1)(w-t+1).\]

\hfill{}$\square$

\section{Doubly Steiner system}

We start this section with new definitions :

A \textbf{$(w_{1},n_{1},w_{2},n_{2},d=w_{1}+w_{2}-t_{1}-t_{2}+1)$}
code is \emph{perfect} $(t_{1},t_{2})$ \emph{cover} if every word
from configuration \textbf{$(t_{1},t_{2})$} is contained in exactly
one codeword. Note, that all the codewords are from configuration
$(w_{1},w_{2})$. The definition of doubly constant weight code which
is a perfect cover is akin to a constant weight code which is a Steiner
system. Hence, one can call such a code \emph{doubly Steiner system}
$S(t_{1},t_{2},w_{1},w_{2},n_{1},n_{2})$.

In \cite{doubly} Etzion show that a doubly Steiner system $S(t_{1},t_{2},w_{1},w_{2},n_{1},n_{2})$
is an optimal $(w_{1},n_{1},w_{2},n_{2},{}_{1}+w_{2}-t_{1}-t_{2}+1))$
code, and present the bounds on the length of such code.

In the follows we prove that the doubly Steiner system is a diameter
perfect code and present the new bound on its length, equivalent to
bound of Tits for Steiner system.

\noindent \textbf{Lemma 65}. Any doubly Steiner system $S(t_{1},t_{2},w_{1},w_{2},n_{1},n_{2})$
forms a diameter perfect code.

\noindent \emph{Proof.} Let $C$ be a $(w_{1},n_{1},w_{2},n_{2},(w_{1}+w_{2}-t_{1}-t_{2}+1))$
code which is a perfect $(t_{1},t_{2})$-cover corresponding to a
$S(t_{1},t_{2},w_{1},w_{2},n_{1},n_{2})$. Then \[
|C|=\frac{\binom{n_{1}}{t_{1}}\binom{n_{2}}{t_{2}}}{\binom{w_{1}}{t_{1}}\binom{w_{2}}{t_{2}}}=\frac{\binom{n_{1}}{w_{1}}\binom{n_{2}}{w_{2}}}{\binom{n_{1}-t_{1}}{w_{1}-t_{1}}\binom{n_{2}-t_{2}}{w_{2}-t_{2}}}.\]
On the other hand by Theorem 62, $|C|\leq\frac{\binom{n_{1}}{w_{1}}\binom{n_{2}}{w_{2}}}{A^{*}(w_{1},n_{1}w_{2},n_{2},(w_{1}+w_{2}-t_{1}-t_{2}))}$,
where $A^{*}(w_{1},n_{1},w_{2},$ $n_{2},(w_{1}+w_{2}-t_{1}-t_{2}))$
is an optimal anticode with diameter $(w_{1}+w_{2}-t_{1}-t_{2})$.
Therefore \[
A^{*}(w_{1},n_{1}w_{2},n_{2},(w_{1}+w_{2}-t_{1}-t_{2}))\leq\binom{n_{1}-t_{1}}{w_{1}-t_{1}}\binom{n_{2}-t_{2}}{w_{2}-t_{2}}.\]
We construct an anticode of size $\binom{n_{1}-t_{1}}{w_{1}-t_{1}}\binom{n_{2}-t_{2}}{w_{2}-t_{2}}$
as follows. We take a constant set of coordinates of size $t_{1}$
in the first $n_{1}$ coordinates and $t_{2}$ in the last $n_{2}$
coordinates and complete it by all vectors of size $w_{1}-t_{1}$
in the first part and $w_{2}-t_{2}$ in the last part .

Since there exists an anticode of size $\binom{n_{1}-t_{1}}{w_{1}-t_{1}}\binom{n_{2}-t_{2}}{w_{2}-t_{2}}$
, the statement follows.

\hfill{}$\square$

\noindent \textbf{Corollary 66}\textbf{\emph{.}}

For any anticode $A(w_{1},n_{1}w_{2},n_{2},(w_{1}+w_{2}-t_{1}-t_{2}))$
with diameter $w_{1}+w_{2}-t_{1}-t_{2}$ (Johnson distance) we have\[
A(w_{1},n_{1}w_{2},n_{2},(w_{1}+w_{2}-t_{1}-t_{2}))\leq\binom{n_{1}-t_{1}}{w_{1}-t_{1}}\binom{n_{2}-t_{2}}{w_{2}-t_{2}}.\]

\noindent \textbf{Theorem 67}\textbf{\emph{.}} If a doubly Steiner
system $S(t_{1},t_{2},w_{1},w_{2},n_{1},n_{2})$ exists and $t_{2}>t_{1},$$t_{1}<w_{1}$
then\[
n_{1}\geq(t_{1}+1)w_{1}-t_{1}t_{2}\]

\[
n_{2}\geq(t_{2}+1)(w_{2}-t_{2}+1).\]

\noindent \emph{Proof.} Let $C$ be a $(w_{1},n_{1},w_{2},n_{2},(w_{1}+w_{2}-t_{1}-t_{2}+1))$
code which is a perfect $(t_{1},t_{2})$-cover, corresponding to a
$S(t_{1},t_{2},w_{1},w_{2},n_{1},n_{2})$.

Let $S$ be a vector from configuration $(t_{1}+1,t_{2}),$ which
is not contained in any codeword. Consider $t_{1}+1$ subvectors of
$S$ from configuration $(t_{1},t_{2}).$ Each of them is contained
in exactly one codeword. Since the minimal distance of the code is
$w_{1}+w_{2}-t_{1}-t_{2}+1$, there are precisely $t_{1}+1$ codewords
which contain those vectors, and these $t_{1}+1$ codewords are disjoint
outside of $S.$ Therefore in the first $n_{1}$ coordinates we have:\[
n_{1}-(t_{1}+1)\geq(w_{1}-t_{1})(t_{1}+1)\]
or \[
n_{1}\geq(w_{1}-t_{1}+1)(t_{1}+1)\]
and in the last $n_{2}$ coordinates we have:\[
n_{2}-t_{2}\geq(w_{2}-t_{2})(t_{1}+1)\]
or \[
n_{2}\geq w_{2}(t_{1}+1)-t_{1}t_{2}.\]
By swapping the roles of $n_{1}$ and $n_{2}$, and $w_{1}$ and $w_{2}$,
we get that

\[
n_{1}\geq w_{1}(t_{2}+1)-t_{1}t_{2}.\]
Therefore, \[
n_{1}\geq\max\{w_{1}(t_{2}+1)-t_{1}t_{2},\:(w_{1}-t_{1}+1)(t_{1}+1)\}\]

\[
n_{2}\geq\max\{w_{2}(t_{1}+1)-t_{1}t_{2},\:(w_{2}-t_{2}+1)(t_{2}+1)\}\]
If we write $t_{2}=t_{1}+a$, where $a>0$ is an integer, we can rewrite
the last expression as follows:

\[
\max\{w_{1}(t_{1}+a+1)-t_{1}(t_{1}+a),\:(w_{1}-t_{1}+1)(t_{1}+1)\}\]
\[
=\max\{w_{1}t_{1}+w_{1}-t_{1}^{2}+(w_{1}-t_{1})a,\: w_{1}t_{1}+w_{1}-t_{1}^{2}+1\}\]

\[
=w_{1}t_{1}+w_{1}-t_{1}^{2}+(w_{1}-t_{1})a\]
Therefore, \[
n_{1}\geq w_{1}(t_{2}+1)-t_{1}t_{2}.\]
Similarly we obtain\[
\max\{w_{2}(t_{1}+1)-t_{1}t_{2},\:(w_{2}-t_{2}+1)(t_{2}+1)\}\]

\[
=(w_{2}-t_{2}+1)(t_{2}+1).\]
Therefore, \[
n_{2}\geq(t_{2}+1)(w_{2}-t_{2}+1).\]

\hfill{}$\square$

\appendix

\chapter*{Appendix A}

\addcontentsline{toc}{chapter}{Appendix A}

\section*{Proof of Proposition 48.}

\textbf{Proposition 48.} For all $k$, $\varphi<k\leq w$, we have
\[
0=[1+k^{2}-k(1+n)+nw-w^{2}]\sum_{i=0}^{w}\binom{i}{k}\Delta_{i}+(1-k+w)^{2}\sum_{i=0}^{w}\binom{i}{k-1}\Delta_{i}\]

\noindent \emph{Proof}\textbf{.} In the Section 2 we saw that for
all $k$, $\varphi<k\leq w$,\begin{eqnarray*}
0 & = & \sum_{i=0}^{w}\binom{i}{k}(i+1)^{2}\Delta_{i+1}+(n-2w)\sum_{i=0}^{w}\binom{i}{k}(i+1)\Delta_{i+1}+\sum_{i=0}^{w}\binom{i}{k}\Delta_{i}\\
 & + & w(n-2w)\sum_{i=0}^{w}\binom{i}{k}\Delta_{i}+(4w-n)\sum_{i=0}^{w}\binom{i}{k}\, i\Delta_{i}-2\sum_{i=0}^{w}\binom{i}{k}\, i^{2}\Delta_{i}\\
 & + & w^{2}\sum_{i=0}^{w}\binom{i}{k}\Delta_{i-1}-2w\sum_{i=0}^{w}\binom{i}{k}(i-1)\Delta_{i-1}+\sum_{i=0}^{w}\binom{i}{k}(i-1)^{2}\Delta_{i-1}\\
\\\end{eqnarray*}
Now we simplify it, using follows equations:\[
\binom{i}{k}(i+1)=\binom{i+1}{k+1}(k+1)\]

\[
\binom{i}{k}=\binom{i-1}{k-1}+\binom{i-1}{k}\]

\[
\binom{i}{k-1}i^{2}+(k+1)\binom{i}{k+1}i-\binom{i}{k}i^{2}=i(k-1)\binom{i}{k-1}\]

\[
\binom{i}{k-1}i=k\binom{i}{k}+(k-1)\binom{i}{k-1}\]

\begin{eqnarray*}
0 & = & (k+1)\sum_{i=0}^{w}\binom{i+1}{k+1}(i+1)\Delta_{i+1}+(n-2w)(k+1)\sum_{i=0}^{w}\binom{i+1}{k+1}\Delta_{i+1}\\
 & + & \sum_{i=0}^{w}\binom{i}{k}\Delta_{i}+w(n-2w)\sum_{i=0}^{w}\binom{i}{k}\Delta_{i}+(4w-n)\sum_{i=0}^{w}\binom{i}{k}\, i\Delta_{i}-2\sum_{i=0}^{w}\binom{i}{k}\, i^{2}\Delta_{i}\\
 & + & w^{2}\sum_{i=0}^{w}\binom{i-1}{k-1}\Delta_{i-1}+w^{2}\sum_{i=0}^{w}\binom{i-1}{k}\Delta_{i-1}-2w\sum_{i=0}^{w}\binom{i-1}{k-1}(i-1)\Delta_{i-1}\\
 & - & 2w\sum_{i=0}^{w}\binom{i-1}{k}(i-1)\Delta_{i-1}+\sum_{i=0}^{w}\binom{i-1}{k-1}(i-1)^{2}\Delta_{i-1}+\sum_{i=0}^{w}\binom{i-1}{k}(i-1)^{2}\Delta_{i-1}\\
\\\end{eqnarray*}

\begin{eqnarray*}
0 & = & (k+1)\sum_{i=0}^{w}\binom{i}{k+1}i\Delta_{i}+(n-2w)(k+1)\sum_{i=0}^{w}\binom{i}{k+1}\Delta_{i}+\sum_{i=0}^{w}\binom{i}{k}\Delta_{i}\\
 & + & w(n-2w)\sum_{i=0}^{w}\binom{i}{k}\Delta_{i}+(4w-n)\sum_{i=0}^{w}\binom{i}{k}\, i\Delta_{i}-2\sum_{i=0}^{w}\binom{i}{k}\, i^{2}\Delta_{i}\\
 & + & w^{2}\sum_{i=0}^{w}\binom{i}{k-1}\Delta_{i}+w^{2}\sum_{i=0}^{w}\binom{i}{k}\Delta_{i}-2w\sum_{i=0}^{w}\binom{i}{k-1}i\Delta_{i}\\
 & - & 2w\sum_{i=0}^{w}\binom{i}{k}i\Delta_{i}+\sum_{i=0}^{w}\binom{i}{k-1}i^{2}\Delta_{i}+\sum_{i=0}^{w}\binom{i}{k}i^{2}\Delta_{i}\end{eqnarray*}

\begin{eqnarray*}
0 & = & \underline{(k+1)\sum_{i=0}^{w}\binom{i}{k+1}i\Delta_{i}}+\underline{\underline{(n-2w)(k+1)\sum_{i=0}^{w}\binom{i}{k+1}\Delta_{i}}}\\
 & + & (wn-w^{2}+1)\sum_{i=0}^{w}\binom{i}{k}\Delta_{i}+\underline{\underline{(2w-n)\sum_{i=0}^{w}\binom{i}{k}\, i\Delta_{i}}}-\underline{\sum_{i=0}^{w}\binom{i}{k}i^{2}\Delta_{i}}\\
 & + & w^{2}\sum_{i=0}^{w}\binom{i}{k-1}\Delta_{i}-2w\sum_{i=0}^{w}\binom{i}{k-1}i\Delta_{i}+\underline{\sum_{i=0}^{w}\binom{i}{k-1}i^{2}\Delta_{i}}\end{eqnarray*}

\begin{eqnarray*}
0 & = & \underline{(k-1)\sum_{i=0}^{w}\binom{i}{k-1}i\Delta_{i}}+\underline{\underline{(2w-n)k\sum_{i=0}^{w}\binom{i}{k}\Delta_{i}}}+(wn-w^{2}+1)\sum_{i=0}^{w}\binom{i}{k}\Delta_{i}\\
 & + & w^{2}\sum_{i=0}^{w}\binom{i}{k-1}\Delta_{i}-2w\sum_{i=0}^{w}\binom{i}{k-1}i\Delta_{i}\\
 & = & (k-1-2w)\sum_{i=0}^{w}\binom{i}{k-1}i\Delta_{i}+w^{2}\sum_{i=0}^{w}\binom{i}{k-1}\Delta_{i}\\
 & + & [(2w-n)k+wn-w^{2}+1]\sum_{i=0}^{w}\binom{i}{k}\Delta_{i}\\
 & = & [(2w-n)k+wn-w^{2}+1+k(k-1-2w)]\sum_{i=0}^{w}\binom{i}{k}\Delta_{i}\\
 & + & [w^{2}+(k-1)(k-1-2w)]\sum_{i=0}^{w}\binom{i}{k-1}\Delta_{i}\end{eqnarray*}
Finally, we get:\[
0=[1+k^{2}-k(1+n)+nw-w^{2}]\sum_{i=0}^{w}\binom{i}{k}\Delta_{i}+(1-k+w)^{2}\sum_{i=0}^{w}\binom{i}{k-1}\Delta_{i}\]

\hfill{}$\square$

\chapter*{Appendix B}

\addcontentsline{toc}{chapter}{Appendix B}

\section*{Proof of Proposition 49}

\textbf{Proposition 49.} For each $k$, $\varphi<k\leq w$, we have\begin{eqnarray*}
0 & = & \frac{1}{4}[4+k^{4}+5w^{2}-2w^{3}+w^{4}-2k^{3}(1+2w)+k^{2}(7+2w+6w^{2})\\
 & - & 2k(3+5w-w^{2}+2w^{3})]\sum_{i=0}^{w}\binom{i}{k}\Delta_{i}\\
 & + & \frac{1}{2}(1-k+w)^{2}(4+k^{2}+w^{2}-2k(1+w))\sum_{i=0}^{w}\binom{i}{k-1}\Delta_{i}\\
 & + & \frac{1}{4}(1-k+w)^{2}(2-k+w)^{2}\sum_{i=0}^{w}\binom{i}{k-2}\Delta_{i}\end{eqnarray*}

\noindent \emph{Proof.} We use the following identities:

\begin{eqnarray*}
\binom{i}{k}(i+1) & = & (k+1)\binom{i+1}{k+1}\\
\binom{i}{k}\binom{i+2}{2} & = & \frac{(k+1)(k+2)}{2}\binom{i+2}{k+2}\\
\binom{i}{k} & = & \binom{i-1}{k-1}+\binom{i-1}{k}=\binom{i-2}{k-2}+2\binom{i-2}{k-1}+\binom{i-2}{k}\end{eqnarray*}
for the calculations below.

\begin{eqnarray*}
0 & = & \frac{(k+1)(k+2)}{2}\sum_{i=0}^{w}\binom{i+2}{k+2}\binom{i+2}{2}\Delta_{i+2}\\
 & + & \sum_{i=0}^{w}\left[\binom{i-2}{k-2}+2\binom{i-2}{k-1}+\binom{i-2}{k}\right]\binom{w-(i-2)}{2}^{2}\Delta_{i-2}\\
 & + & (k+1)\sum_{i=0}^{w}\binom{i+1}{k+1}\left[(i+1)+2(w-(i+1))\binom{i+1}{2}\right]\Delta_{i+1}\\
 & + & \sum_{i=0}^{w}\left[\binom{i-1}{k-1}+\binom{i-1}{k}\right]\\
 & * & \left[(w-(i-1)^{2}+2(i-1)(w-(i-1))\binom{w-(i-1)}{2}\right]\Delta_{i-1}\\
 & + & \sum_{i=0}^{w}\binom{i}{k}\left[1+2i(w-i)+2\binom{i}{2}\binom{w-i}{2}+i^{2}(w-i)^{2}\right]\Delta_{i}\end{eqnarray*}

\begin{eqnarray*}
0 & = & \frac{(k+1)(k+2)}{2}\sum_{i=0}^{w}\binom{i}{k+2}\binom{i}{2}\Delta_{i}\\
 & + & \sum_{i=0}^{w}\left[\binom{i}{k-2}+2\binom{i}{k-1}+\binom{i}{k}\right]\binom{w-i}{2}^{2}\Delta_{i}\\
 & + & (k+1)\sum_{i=0}^{w}\binom{i}{k+1}\left[i+2(w-i)\binom{i}{2}\right]\Delta_{i}\\
 & + & \sum_{i=0}^{w}\left[\binom{i}{k-1}+\binom{i}{k}\right]\left[(w-i)^{2}+2i(w-i)\binom{w-i)}{2}\right]\Delta_{i}\\
 & + & \sum_{i=0}^{w}\binom{i}{k}\left[1+2i(w-i)+2\binom{i}{2}\binom{w-i}{2}+i^{2}(w-i)^{2}\right]\Delta_{i}\end{eqnarray*}

\begin{eqnarray*}
0 & = & \frac{(k+1)(k+2)}{4}\sum_{i=0}^{w}\binom{i}{k+2}(i^{2}-i)\Delta_{i}\\
 & + & \frac{1}{4}\sum_{i=0}^{w}\binom{i}{k-2}(w-i)^{2}(w-i-1)^{2}\Delta_{i}\\
 & + & \sum_{i=0}^{w}\binom{i}{k-1}\left[\frac{(w-i)^{2}(w-i-1)^{2}}{2}+\left[(w-i)^{2}+i(w-i)^{2}(w-i-1)\right]\right]\Delta_{i}\\
 & + & (k+1)\sum_{i=0}^{w}\binom{i}{k+1}\left[i+(w-i)(i-1)i\right]\Delta_{i}\\
 & + & \sum_{i=0}^{w}\binom{i}{k}\left[\frac{(w-i)^{2}(w-i-1)^{2}}{4}+(w-i)^{2}+i(w-i)^{2}(w-i-1)+1\right.\\
 & + & \left.2i(w-i)+\frac{i(i-1)(w-i)(w-i-1)}{2}+i^{2}(w-i)^{2}\right]\Delta_{i}\end{eqnarray*}
Finally we obtain:\begin{eqnarray*}
0 & = & \frac{1}{4}(k+1)(k+2)\sum_{i=0}^{w}\binom{i}{k+2}[i^{2}-i]\Delta_{i}\\
 & + & (k+1)\sum_{i=0}^{w}\binom{i}{k+1}\left[(1-w)i+(1+w)i^{2}-i^{3}\right]\Delta_{i}\\
 & + & \sum_{i=0}^{w}\binom{i}{k}\left[(1+\frac{5w^{2}}{4}-\frac{w^{3}}{2}+\frac{w^{4}}{4})+(-\frac{5}{4}+w)i^{2}+(-\frac{1}{2}-w)i^{3}+\frac{3}{4}i^{4}\right]\Delta_{i}\\
 & + & \sum_{i=0}^{w}\binom{i}{k-1}\left[(\frac{3}{2}w^{2}-w^{3}+\frac{w^{4}}{2})+(-3w+2w^{2}-w^{3})i\right.\\
 & + & \left.(\frac{3}{2}-w)i^{2}+wi^{3}-\frac{1}{2}i^{4}\right]\Delta_{i}\\
 & + & \frac{1}{4}\sum_{i=0}^{w}\binom{i}{k-2}\left[(w^{2}-2w^{3}+w^{4})+(-2w+6w^{2}-4w^{3})i\right.\\
 & + & \left.(1-6w+6w^{2})i^{2}+(2-4w)i^{3}+i^{4}\right]\Delta_{i}\end{eqnarray*}

Now we use following identities for computation of coefficients of
$\sum_{i=0}^{w}\binom{i}{j}$ for $j=k-2,...,k+4$:

\[
\binom{i}{k}\, i=(k+1)\binom{i}{k+1}+k\binom{i}{k}\]

\[
\binom{i}{k}\, i^{2}=(k+1)(k+2)\binom{i}{k+2}+(k+1)(2k+1)\binom{i}{k+1}+k^{2}\binom{i}{k}\]
\begin{eqnarray*}
\binom{i}{k}\, i^{3} & = & (k+1)(k+2)(k+3)\binom{i}{k+3}+3(k+1)^{2}(k+2)\binom{i}{k+2}\\
 & + & (k+1)(3k^{2}+3k+1)\binom{i}{k+1}+k^{3}\binom{i}{k}\end{eqnarray*}

\begin{eqnarray*}
\binom{i}{k}\, i^{4} & = & (k+1)(k+2)(k+3)(k+4)\binom{i}{k+4}\\
 & + & (4k+6)(k+1)(k+2)(k+3)\binom{i}{k+3}\\
 & + & (k+1)(k+2)(6k^{2}+12k+7)\binom{i}{k+2}\\
 & + & (k+1)(4k^{3}+6k^{2}+4k+1)\binom{i}{k+1}+k^{4}\binom{i}{k}\end{eqnarray*}

The coefficient of $\sum_{i=0}^{w}\binom{i}{k+4}\Delta_{i}$ follows
from $\binom{i}{k}i^{4}$, $\binom{i}{k+1}i^{3}$, $\binom{i}{k+2}i^{2}$,
therefore it equals to:

\begin{eqnarray*}
 &  & \frac{1}{4}(k+1)(k+2)(k+3)(k+4)+(k+1)(-(k+2)(k+3)(k+4))\\
 & + & \frac{3}{4}(k+1)(k+2)(k+3)(k+4)\\
 & = & (k+1)(k+2)(k+3)(k+4)[\frac{1}{4}-1+\frac{3}{4}]=0\\
\\\end{eqnarray*}

The coefficient of $\sum_{i=0}^{w}\binom{i}{k+3}\Delta_{i}$ follows
from $\binom{i}{k+2}i^{2}$, $\binom{i}{k+2}i$, $\binom{i}{k+1}i^{2}$,
$\binom{i}{k+1}i^{3}$, $\binom{i}{k}i^{3}$, $\binom{i}{k}i^{4}$,
$\binom{i}{k-1}i^{4}$, therefore it equals to:

\begin{eqnarray*}
 &  & \frac{(k+1)(k+2)}{4}(k+3)(2k+5)-\frac{(k+1)(k+2)}{4}(k+3)\\
 & + & (k+1)(1+w)(k+2)(k+3)\\
 & - & (k+1)3(k+2)^{2}(k+3)+(-\frac{1}{2}-w)(k+1)(k+2)(k+3)\\
 & + & \frac{3}{4}(4k+6)(k+1)(k+2)(k+3)-\frac{1}{2}k(k+1)(k+2)(k+3)\\
 & = & (k+1)(k+2)(k+3)\left[\frac{2k+5}{4}-\frac{1}{4}+(1+w)-3(k+2)\right.\\
 & + & \left.(-\frac{1}{2}-w)+\frac{3}{4}(4k+6)-\frac{1}{2}k\right]\\
 & = & (k+1)(k+2)(k+3)*0=0\\
\\\end{eqnarray*}
The coefficient of $\sum_{i=0}^{w}\binom{i}{k+2}\Delta_{i}$ follows
from $\binom{i}{k+2}i^{2}$, $\binom{i}{k+2}i$, $\binom{i}{k+1}i$,
$\binom{i}{k+1}i^{2}$, $\binom{i}{k+1}i^{3}$, $\binom{i}{k}i^{2}$,
$\binom{i}{k}i^{3}$, $\binom{i}{k}i^{4}$, $\binom{i}{k-1}i^{3}$,
$\binom{i}{k-1}i^{4}$, $\binom{i}{k-2}i^{4}$, therefore it equals
to:\begin{eqnarray*}
 &  & \frac{(k+1)(k+2)}{4}(k+2)^{2}-\frac{(k+1)(k+2)}{4}(k+2)+(k+1)(1-w)(k+2)\\
 & + & (k+1)(1+w)(k+2)(2k+3)-(k+1)(k+2)(3(k+1)^{2}\\
 & + & 3(k+1)+1)+(-\frac{5}{4}+w)(k+1)(k+2)+(-\frac{1}{2}-w)3(k+1)^{2}(k+2)\\
 & + & \frac{3}{4}(k+1)(k+2)(6k^{2}+12k+7)+wk(k+1)(k+2)\\
 & - & \frac{1}{2}(4(k-1)+6)k(k+1)(k+2)+\frac{1}{4}(k-1)k(k+1)(k+2)\\
 & = & (k+1)(k+2)\left[\frac{(k+2)^{2}}{4}-\frac{k+2}{4}+(1-w)+(1+w)(2k+3)\right.\\
 & - & (3(k+1)^{2}+3k+4)+(-\frac{5}{4}+w)(-\frac{1}{2}-w)3(k+1)\\
 & + & \left.\frac{3}{4}(6k^{2}+12k+7)+wk-\frac{1}{2}(4k+2)k+\frac{1}{4}(k-1)k\right]\\
 & = & (k+1)(k+2)*0=0\\
\\\end{eqnarray*}
The coefficient of $\sum_{i=0}^{w}\binom{i}{k+1}\Delta_{i}$ follows
from $\binom{i}{k+1}i$, $\binom{i}{k+1}i^{2}$, $\binom{i}{k+1}i^{3}$,
$\binom{i}{k}i^{2}$, $\binom{i}{k}i^{3}$, $\binom{i}{k}i^{4}$,
$\binom{i}{k-1}i^{2}$, $\binom{i}{k-1}i^{3}$, $\binom{i}{k-1}i^{4}$,
$\binom{i}{k-2}i^{3}$, $\binom{i}{k-2}i^{4}$, therefore it equals
to:\begin{eqnarray*}
 &  & (k+1)(1-w)(k+1)+(k+1)(1+w)(k+1)^{2}-(k+1)(k+1)^{3}\\
 & + & (-\frac{5}{4}+w)(k+1)(2k+1)+(-\frac{1}{2}-w)(k+1)(3k^{2}+3k+1)\\
 & + & (\frac{3}{4}(k+1)(4k^{3}+6k^{2}+4k+1)+(\frac{3}{2}-w)k(k+1)+w3k^{2}(k+1)\\
 & - & \frac{1}{2}k(k+1)(6(k-1)^{2}+12(k-1)+7)\\
 & + & \frac{(2-4w)}{4}(k-1)k(k+1)+\frac{1}{4}(4(k-2)+6)(k-1)k(k+1)=0\\
\\\end{eqnarray*}
The coefficient of $\sum_{i=0}^{w}\binom{i}{k}\Delta_{i}$ follows
from $\binom{i}{k}$, $\binom{i}{k}i^{2}$, $\binom{i}{k}i^{3}$,
$\binom{i}{k}i^{4}$, $\binom{i}{k-1}i$, $\binom{i}{k-1}i^{2}$,
$\binom{i}{k-1}i^{3}$, $\binom{i}{k-1}i^{4}$, $\binom{i}{k-2}i^{2}$,
$\binom{i}{k-2}i^{3}$, $\binom{i}{k-2}i^{4}$, therefore it equals
to: \begin{eqnarray*}
 &  & (1+\frac{5w^{2}}{4}-\frac{w^{3}}{2}+\frac{w^{4}}{4})+(-\frac{5}{4}+w)k^{2}+(-\frac{1}{2}-w)k^{3}+\frac{3}{4}k^{4}+(-3w+2w^{2}-w^{3})k\\
 & + & (\frac{3}{2}-w)k(2(k-1)+1)+wk(3(k-1)^{2}+3(k-1)+1)\\
 & - & \frac{1}{2}k(4(k-1)^{3}+6(k-1)^{2}+4(k-1)+1)+\frac{(1-6w+6w^{2})}{4}(k-1)k\\
 & + & \frac{(2-4w)}{4}3(k-1)^{2}k+\frac{1}{4}(k-1)k(6(k-2)^{2}+12(k-2)+7)\\
 & = & \frac{1}{4}(4+k^{4}+5w^{2}-2w^{3}+w^{4}-2k^{3}(1+2w)+k^{2}(7+2w+6w^{2})\\
 & - & 2k(3+5w-w^{2}+2w^{3}))\end{eqnarray*}
Now we calculate the two remaining coefficients:

\noindent The coefficient of $\sum_{i=0}^{w}\binom{i}{k-1}\Delta_{i}$
follows from $\binom{i}{k-1}$, $\binom{i}{k-1}i$, $\binom{i}{k-1}i^{2}$,
$\binom{i}{k-1}i^{3}$, $\binom{i}{k-1}i^{4}$, $\binom{i}{k-2}i$,
$\binom{i}{k-2}i^{2}$, $\binom{i}{k-2}i^{3}$, $\binom{i}{k-2}i^{4}$,
therefore it equals to: \begin{eqnarray*}
 &  & (\frac{3}{2}w^{2}-w^{3}+\frac{w^{4}}{2})+(-3w+2w^{2}-w^{3})(k-1)+(\frac{3}{2}-w)(k-1)^{2}+w(k-1)^{3}\\
 & - & \frac{1}{2}(k-1)^{4}+\frac{1}{4}(-2w+6w^{2}-4w^{3})(k-1)\\
 & + & \frac{1}{4}(1-6w+6w^{2})(k-1)(2(k-2)+1)+\frac{1}{4}(2-4w)(k-1)(3(k-2)^{2}\\
 & + & 3(k-2)+1)+\frac{1}{4}(k-1)(4(k-2)^{3}+6(k-2)^{2}+4(k-2)+1)\\
 & = & \frac{1}{2}(1-k+w)^{2}(4+k^{2}+w^{2}-2k(1+w))\\
\\\end{eqnarray*}
 The coefficient of $\sum_{i=0}^{w}\binom{i}{k-2}\Delta_{i}$ follows
from $\binom{i}{k-2}$, $\binom{i}{k-2}i$, $\binom{i}{k-2}i^{2}$,
$\binom{i}{k-2}i^{3}$, $\binom{i}{k-2}i^{4}$, therefore it equals
to:\begin{eqnarray*}
 &  & \frac{1}{4}(w^{2}-2w^{3}+w^{4})+\frac{1}{4}(-2w+6w^{2}-4w^{3})(k-2)+\frac{1}{4}(1-6w+6w^{2})(k-2)^{2}\\
 & + & \frac{1}{4}(2-4w)(k-3)^{3}+(k-2)^{4}\\
 & = & \frac{1}{4}(1-k+w)^{2}(2-k+w)^{2}\end{eqnarray*}
Finally, we get the following formula:\begin{eqnarray*}
0 & = & \frac{1}{4}[4+k^{4}+5w^{2}-2w^{3}+w^{4}-2k^{3}(1+2w)+k^{2}(7+2w+6w^{2})\\
 & - & 2k(3+5w-w^{2}+2w^{3})]\sum_{i=0}^{w}\binom{i}{k}\Delta_{i}\\
 & + & \frac{1}{2}(1-k+w)^{2}(4+k^{2}+w^{2}-2k(1+w))\sum_{i=0}^{w}\binom{i}{k-1}\Delta_{i}\\
 & + & \frac{1}{4}(1-k+w)^{2}(2-k+w)^{2}\sum_{i=0}^{w}\binom{i}{k-2}\Delta_{i}\end{eqnarray*}

\hfill{}$\square$

\chapter*{Appendix C}

\addcontentsline{toc}{chapter}{Appendix C}

\section*{Proof of proposition 52}

\textbf{Proposition} \textbf{52.}\[
\sum_{i=0}^{w}[\binom{w-i-1}{k-1}(w-i)^{2}-\binom{w-i}{k-1}i^{2}+\binom{i-1}{k-1}i^{2}-\binom{i}{k-1}(w-i)^{2}]A_{i}=\]
\[
=2(2wk-k^{2}+k)\sum_{i=0}^{w}\binom{i}{k}A_{i}-2(w-(k-1))^{2}\sum_{i=0}^{w}\binom{i}{k-1}A_{i}.\]
\textbf{Proof.} The code is self-complement by \cite{nonexist}, thus
$A_{i}=A_{w-i}$, therefore

\begin{eqnarray*}
2w^{2}\sum_{i=0}^{w}\binom{i}{k}\Delta_{i} & = & \sum_{i=0}^{w}\binom{w-i-1}{k-1}(w-i)^{2}A_{w-i}-\sum_{i=0}^{w}\binom{w-i}{k-1}i^{2}A_{w-i}\\
 & + & \sum_{i=0}^{w}\binom{i-1}{k-1}i^{2}A_{i}-\sum_{i=0}^{w}\binom{i}{k-1}(w-i)^{2}A_{i}\\
 & = & \sum_{i=0}^{w}\binom{i-1}{k-1}i^{2}A_{i}-\sum_{i=0}^{w}\binom{i}{k-1}(w-i)^{2}A_{i}\\
 & + & \sum_{i=0}^{w}\binom{i-1}{k-1}i^{2}A_{i}-\sum_{i=0}^{w}\binom{i}{k-1}(w-i)^{2}A_{i}\\
 & = & 2\sum_{i=0}^{w}\binom{i-1}{k-1}i^{2}A_{i}-2\sum_{i=0}^{w}\binom{i}{k-1}(w-i)^{2}A_{i}\end{eqnarray*}
now we use the following equalities:\[
\binom{i-1}{k-1}i^{2}=\binom{i}{k}ik\]

\[
\binom{i}{k}ik-\binom{i}{k-1}i^{2}=-i(k-1)\binom{i}{k-1}\]

\[
i\binom{i}{k-1}=k\binom{i}{k}+(k-1)\binom{i}{k-1}\]

\begin{eqnarray*}
w^{2}\sum_{i=0}^{w}\binom{i}{k}\Delta_{i} & = & \sum_{i=0}^{w}\binom{i}{k}ikA_{i}-w^{2}\sum_{i=0}^{w}\binom{i}{k-1}A_{i}\\
 & + & 2w\sum_{i=0}^{w}\binom{i}{k-1}iA_{i}-\sum_{i=0}^{w}\binom{i}{k-1}i^{2}A_{i}\\
 & = & -(k-1)\sum_{i=0}^{w}\binom{i}{k-1}iA_{i}-w^{2}\sum_{i=0}^{w}\binom{i}{k-1}A_{i}+2w\sum_{i=0}^{w}\binom{i}{k-1}iA_{i}\\
 & = & (2w-(k-1))k\sum_{i=0}^{w}\binom{i}{k}A_{i}\\
 & + & (2w-(k-1))(k-1)\sum_{i=0}^{w}\binom{i}{k-1}A_{i}-w\sum_{i=0}^{w}\binom{i}{k-1}A_{i}\\
 & = & (2wk-k^{2}+k)\sum_{i=0}^{w}\binom{i}{k}A_{i}\\
 & + & (2w(k-1)-(k-1)^{2}-w^{2})\sum_{i=0}^{w}\binom{i}{k-1}A_{i}\\
 & = & (2wk-k^{2}+k)\sum_{i=0}^{w}\binom{i}{k}A_{i}-(w-(k-1))^{2}\sum_{i=0}^{w}\binom{i}{k-1}A_{i}\end{eqnarray*}
Finally, we get\begin{eqnarray*}
w^{2}\sum_{i=0}^{w}\binom{i}{k}\Delta_{i} & = & (2wk-k^{2}+k)\sum_{i=0}^{w}\binom{i}{k}A_{i}-(w-k+1)^{2}\sum_{i=0}^{w}\binom{i}{k-1}A_{i}\end{eqnarray*}

\hfill{}$\square$

\bibliographystyle{is-unsrt}
\bibliography{master}

\addcontentsline{toc}{chapter}{Bibliography}
\end{document}